\documentclass[sigconf,screen]{acmart}

\AtBeginDocument{%
  \providecommand\BibTeX{{%
    \normalfont B\kern-0.5em{\scshape i\kern-0.25em b}\kern-0.8em\TeX}}}

\copyrightyear{2020} 
\acmYear{2020} 
\setcopyright{acmcopyright}
\acmConference[ESEC/FSE '20]{Proceedings of the 28th ACM Joint European Software Engineering Conference and Symposium on the Foundations of Software Engineering}{November 8--13, 2020}{Virtual Event, USA}
\acmBooktitle{Proceedings of the 28th ACM Joint European Software Engineering Conference and Symposium on the Foundations of Software Engineering (ESEC/FSE '20), November 8--13, 2020, Virtual Event, USA}
\acmPrice{15.00}
\acmDOI{10.1145/3368089.3409708}
\acmISBN{978-1-4503-7043-1/20/11}

\usepackage{setspace}
\usepackage{algorithmic}
\usepackage{graphicx}
\usepackage{textcomp}
\usepackage{xcolor}
\usepackage{soul}
\usepackage{comment}
\usepackage[linesnumbered,ruled,vlined]{algorithm2e}
\usepackage{balance}
\usepackage{microtype}
\usepackage{multirow}
\usepackage[normalem]{ulem}
\usepackage{soul}
\usepackage{lipsum}
\usepackage{wrapfig}
\usepackage{graphicx}

\useunder{\uline}{\ul}{}

\usepackage{tikz}
\newcommand*\circled[1]{\tikz[baseline=(char.base)]{\small{\textbf{
			\node[shape=circle,fill,inner sep=0.75pt] (char) {\textcolor{white}{#1}};}}}}

\newcommand{\framework}{\emph{{FrUITeR}}\xspace}
\newcommand{\appflow}{AppFlow}
\newcommand{\guitest}{GTM}
\newcommand{\craft}{CraftDroid}

\definecolor{amethyst}{rgb}{0.6, 0.4, 0.8}
\newcommand{\yixue}[1]{{\color{amethyst}(Yixue: #1)}}

\begin{document}

\title{\framework: A Framework for Evaluating UI Test Reuse}

\author{Yixue Zhao}
\email{yixue.zhao@usc.edu}
\affiliation{%
  \institution{University of Southern California}
  \country{USA}
}
\author{Justin Chen}
\email{justin.chen@columbia.edu}
\affiliation{%
  \institution{Columbia University}
  \country{USA}
}
\author{Adriana Sejfia}
\email{sejfia@usc.edu}
\affiliation{%
  \institution{University of Southern California}
  \country{USA}
}
\author{Marcelo Schmitt Laser}
\email{marcelo.laser@gmail.com}
\affiliation{%
  \institution{University of Southern California}
  \country{USA}
}
\author{Jie Zhang}
\email{jie.zhang@ucl.ac.uk}
\affiliation{
  \institution{University College London}
  \country{UK}
}

\author{Federica Sarro}
\email{f.sarro@ucl.ac.uk}
\affiliation{
  \institution{University College London}
  \country{UK}
}
\author{Mark Harman}
\email{mark.harman@ucl.ac.uk}
\affiliation{
  \institution{University College London}
  \country{UK}
}
\author{Nenad Medvidovic}
\email{neno@usc.edu}
\affiliation{
  \institution{ University of Southern California}
  \country{USA}
}

\renewcommand{\shortauthors}{Y. Zhao, J. Chen, A. Sejfia, M. Laser, J. Zhang, F. Sarro, M. Harman, and N. Medvidovic}

\begin{abstract}
UI testing is tedious and time-consuming due to  the manual effort required.
Recent research has explored  opportunities for reusing existing UI tests from an app to automatically generate new tests for other apps.
However, the evaluation of such techniques currently remains manual, unscalable, and unreproducible, which can waste effort and impede progress in this emerging area.
We introduce \framework, a framework that automatically evaluates UI test reuse in a reproducible way.
We apply \framework~ to existing test-reuse techniques on a uniform benchmark we established, resulting in 11,917 test reuse cases from 20 apps.
We report several key findings aimed at improving UI test reuse that are missed by existing work.
\end{abstract}



\maketitle

\section{introduction}
\label{sec:intro}

Writing UI tests is tedious and time-consuming~\cite{hu2018appflow,linares2017developers}, increasingly driving the focus toward 
automated UI testing~\cite{choudhary2015orsosurvey}. 
However, existing work tends to target tests that yield high code coverage, rather than  \emph{usage-based} tests that
explore an app's functionality, e.g., \emph{sign-in}, \emph{purchase}, \emph{search}, etc.
Developers heavily rely on  usage-based tests \cite{linares2017developers}, but currently have to write them manually~\cite{linares2017developers,choudhary2015orsosurvey}.

To reduce the manual effort of writing usage-based tests, recent research has explored reusing existing  tests in a \emph{source app} to generate new tests automatically for a \emph{target app}~\cite{behrang2018test,zeller2018transferring,hu2018appflow, lin2019craftdroid, behrang2019atm}. 
The guiding insight 
is that different apps expose common functionalities via semantically similar GUI elements.
This  suggests that it is possible to reuse existing UI tests across apps---in effect generating the tests \emph{automatically}---by \emph{mapping} similar GUI elements.

Four recent techniques have targeted usage-based test reuse across Android apps \cite{behrang2018test,hu2018appflow, lin2019craftdroid, behrang2019atm}.\footnote{Rau el al. recently proposed a test-reuse technique for web applications \cite{zeller2018transferring}. In this paper, we focus on Android apps due to the availability of a larger number of existing techniques to evaluate, although in principle our work is not limited to Android.}
While these techniques have shown promise, we have identified five important limitations 
that hinder their comparability, reproducibility, and reusability.
In turn, this can lead to duplication and wasted effort in this emerging area. 

\circled{1} The metrics applied to date evaluate whether GUI events from a source app  are correctly transferred to a target app, but do not consider \emph{whether the transferred tests are actually useful}. 
It is possible that  events are transferred correctly, but the generated test is ``wrong''. This can be, e.g., because a generated test is missing events and thus not executable.
Moreover, the metrics used in existing work are not standardized even when evaluating same aspects of different techniques, making it difficult to compare the techniques. 

\circled{2} Each existing technique's \emph{evaluation process requires significant manual effort}:  every transferred event in each test must be inspected to determine whether the transfer is performed correctly.
This imposes a practical limit on the number of tests that can be evaluated.
For instance, the authors of ATM \cite{behrang2019atm} had to restrict their comparison with \guitest~\cite{behrang2018test} to a randomly selected 50\% of the possible source-target app combinations due to the task's scale. 

\circled{3} There are \emph{no standardized guidelines for conducting the manual inspections}, making the evaluation results biased and hard to reproduce. For instance, ATM's authors acknowledge the possibility of mistakes in the manual process~\cite{behrang2019atm}. Such mistakes are currently hard to locate, verify, or eliminate by other researchers.
\begin{wrapfigure}[15]{r}{0.26\columnwidth}
    \centering
    \includegraphics[width=0.3\columnwidth]{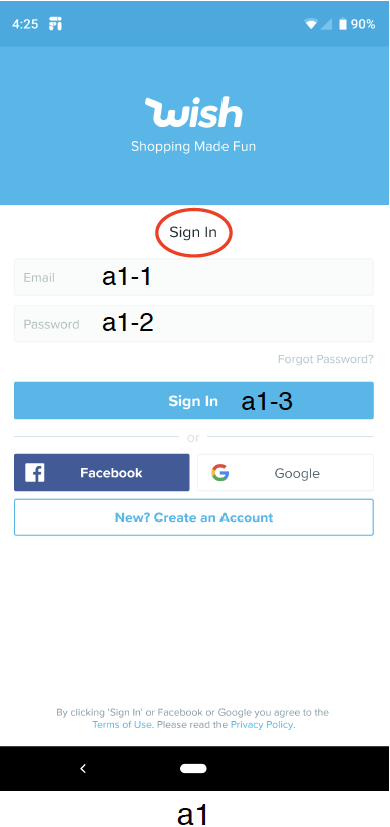}
\end{wrapfigure}
\indent \circled{4} 
Existing techniques are \emph{designed as one-off solutions and evaluated as a whole}.  
This makes  it difficult to isolate and compare their relevant components.
For instance, \guitest~\cite{behrang2018test}, ATM~\cite{behrang2019atm}, and CraftDroid \cite{lin2019craftdroid} all contain functionality to compute a ``similarity score'' 
between two GUI elements, 
 but it is unclear which of those specific components performs the best 
against the same baseline. 
This would impede  subsequent research that could potentially benefit from identifying underlying  components that should be reused and/or improved.
\linebreak \indent
\circled{5} Existing techniques make \emph{different assumptions that hinder their comparison}. 
For instance, \guitest~\cite{behrang2018test} and ATM~\cite{behrang2019atm} require access to  apps' code, and cannot be directly compared with techniques evaluated on close-sourced apps. 

To address limitations 1--3, as well as limitation 4 in part, 
 we have developed \framework, a \textbf{\textit{Fr}}ame\-work for evaluating \textbf{\textit{UI}} \textbf{\textit{Te}}st \textbf{\textit{R}}euse.
 \framework 
  consists of three key elements: a set of \emph{new evaluation metrics} that consolidate the metrics used by existing techniques and expand them to measure important aspects that are currently missed; 
  \emph{two baseline UI test-reuse techniques} that establish the lower- and upper-bounds for the evaluation metrics; and \emph{an~automated workflow} that \emph{modularizes} UI test-reuse functionality and significantly reduces the  manual effort. 
With \framework, one can automatically evaluate test-reuse techniques on apps/tests of interest against the same baseline, thus opening the possibility of large-scale studies. 

To fully address limitation 4, as well as limitation 5, we have extracted the core components from existing techniques and established a benchmark for evaluating and comparing them. Our benchmark currently contains 20 subject apps with 239 test cases, involving 1,082 GUI events. This benchmark is used by \framework~ to evaluate side-by-side the extracted components and the two  baseline components we developed, yielding 11,917 test-reuse instances.

The results obtained by \framework revealed several important findings. 
For example, we have been able to pinpoint specific trade-offs between ML-based (e.g., \appflow) and similarity-based (e.g., ATM) techniques. We have also identified scenarios that may seem counter-intuitive, such as the fact that manually writing tests requires less effort than attempting automated transfer in cer\-tain cases. Finally, performing evaluations on a much larger data corpus allowed us to refute some conclusions reached in prior work.

This paper makes the following contributions.		
	\circled{1}~We develop \framework~ to automatically evaluate UI test reuse with an expanded set of  metrics as compared to existing work, and two baseline techniques that help to provide the lower- and upper-bounds of UI test reuse in a given scenario.
	~\circled{2}~We identify and extract the core components from existing test-reuse techniques,
	 enabling their fair 
comparison. 
	\circled{3}~We establish a reusable benchmark with standardized ground truths that facilitates the reproducibility of  UI test-reuse techniques' evaluation and comparison. 	
	\circled{4}~We use \framework~ to conduct a side-by-side evaluation of the state-of-the-art test-reuse techniques, uncovering several needed improvements in this area. 
	\circled{5}~We make \framework's implementation and all data artifacts  publicly available \cite{fruiterrepo}, directly fostering future research. 

Section~2 introduces a representative example,  terminology, and related work.
Section~3 describes \framework's  requirements and Section 4 its design, 
followed by \framework's instantiation in Section~5. 
Section~6 
discusses our key findings.
Section~7 concludes the paper. 

\section{Background and Related Work}
\label{sec:bg}

In this section, we introduce a motivating example and relevant terminology,  followed by an overview of the strategies pursued by existing work and how they have been evaluated to date.
\begin{figure}[t]
    \vspace{1.3mm}
    \centering
    \includegraphics[width=0.475\textwidth]{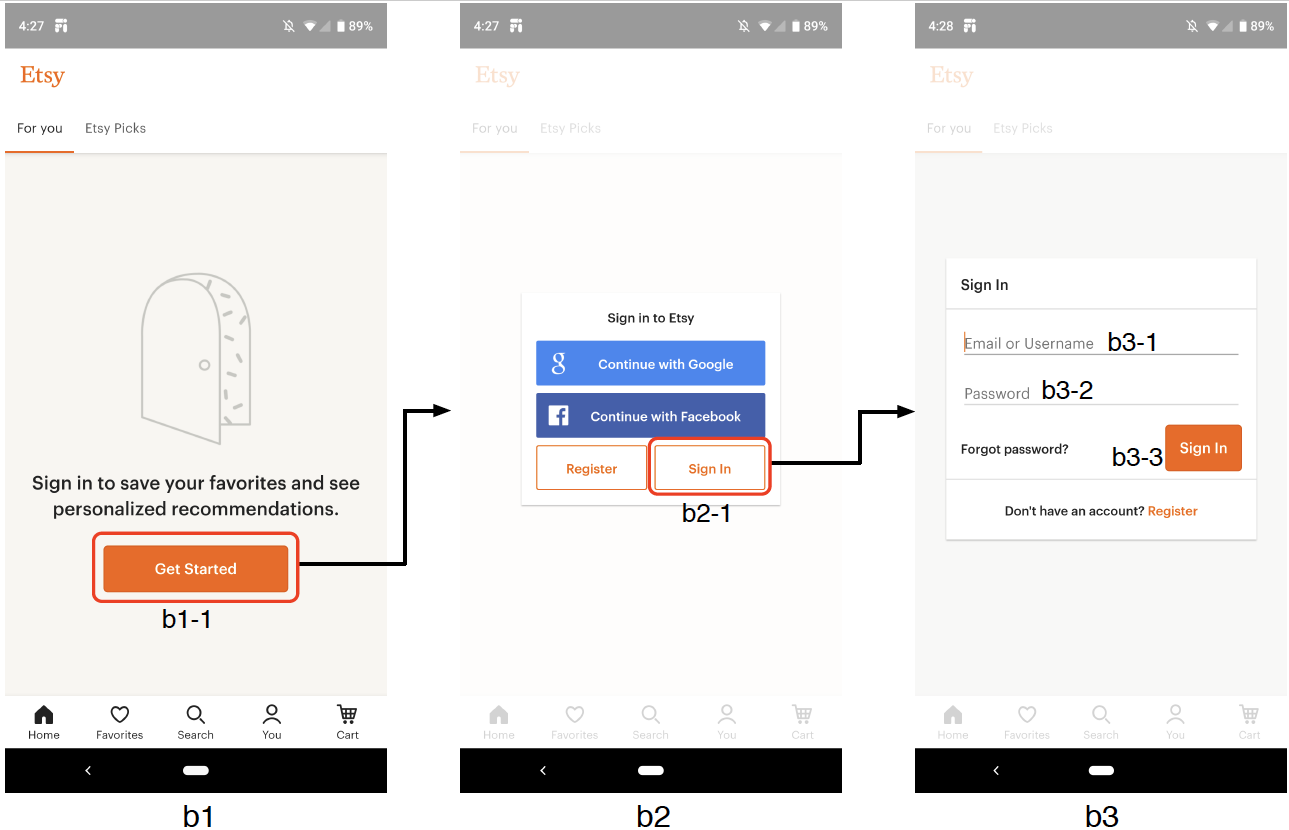}
    \vspace{-5mm}
    \caption{\emph{sign-in} tests for Wish (a1) and Etsy (b1--b3).} 
    \label{fig:example}
    \vspace{-3mm}
\end{figure}
\subsection{Motivating Example and Terminology}
\label{sec:bg:example_term}

Figure~\ref{fig:example} shows the screenshots of the \emph{sign-in} process of two popular shopping apps: Wish (left) and Etsy (right). 
Each screen is labeled with an identifier, e.g., \textsf{a1} is the first screen of Wish. 
In each screen, there may be one or more 
\emph{actionable} GUI elements with which  end-users can interact  based on the associated actions. 
For instance, the ``\textsf{Sign~In}'' button in screen \textsf{a1} (\textsf{a1-3}) is associated with a \textsf{click} action. Actionable elements and their associated actions embody GUI events (defined below).
By contrast, the label ``\textsf{Sign In}'' that is circled in screen \textsf{a1} is a 
\emph{non-actionable} GUI element.

As an illustration, assume that Wish's \emph{sign-in} test exists and our goal is to automatically transfer it to Etsy. 
The relevant actionable GUI elements in this \emph{sign-in} example are labeled and will be used
to describe the following key terms used throughout the paper.

\textbf{GUI Event}, or event in short,
is a triple comprising (1) an \emph{actionable} GUI element, (2) an associated action, and (3)  an optional input value (e.g., user input for a text box). We reuse this  definition from existing work~\cite{behrang2018test, lin2019craftdroid, behrang2019atm}.
For simplicity, we use the label of a GUI element (e.g., \textsf{a1-1}) to refer to the GUI event triple.

\textbf{Canonical Event} is an abstracted event that captures a category of commonly occurring events. An example canonical event may be \textsf{AppSignIn}, and it would correspond to the \textsf{a1-3} and \textsf{b3-3} from Figure \ref{fig:example}, as well as similar events from other apps, such as \textsf{Log In}. 

\textbf{Usage-Based Test}
exercises a given  functionality in an app, such as \emph{sign-in}.
A usage-based test
\footnote{If not mentioned otherwise, test or test case refers to usage-based test in this paper.} 
consists of a sequence of GUI events. For instance, Figure \ref{fig:example} highlights the \emph{sign-in} test in Wish (left) as the event sequence \{ \textsf{a1-1}, \textsf{a1-2}, \textsf{a1-3} \}.

\textbf{Source App} 
is the app with known tests that can be transferred to other apps with similar usage.
For instance, Wish is a source app with a  \emph{sign-in} test that can potentially be transferred  to other apps with \emph{sign-in} functionality.
\textbf{Target App}
is the app to which one aims to transfer existing tests. 
A target app can reuse the tests from multiple source apps; at the same time, it can serve as a source app to other target apps if it contains known tests.
Both source apps and target apps are used extensively in existing work~\cite{behrang2018test, lin2019craftdroid, behrang2019atm}.

\textbf{Source Test} is an existing test for a given source app that should be transferred to a target app to generate a \textbf{Transferred Test}.
\textbf{Ground-Truth Test} is an existing test for a target app that 
is used to evaluate whether the transferred test is correct. (i.e., whether the two tests match).
\textbf{Source Event}, \textbf{Transferred Event}, and \textbf{Ground-Truth Event} refer to the GUI events that belong to the source test, transferred test, and ground-truth test, respectively.

\textbf{Ancillary Event} is a special type of transferred event that is not  mapped from a source event, but is added in order to reach certain states in the target app.
For example, \textsf{b1-1} and \textsf{b2-1} from Figure~\ref{fig:example} may  need to be added as ancillary events 
in order to reach Etsy's \emph{sign-in} screen \textsf{b3}; such events do not exist in the source test. 

\textbf{Null Event} is an event that should have been mapped from a source event, but was not identified as such by a given test-reuse technique. Thus, the null event does not exist in the transferred test, but it has a corresponding source event from which it maps. This could be because of (1) a test-reuse technique's inaccuracy or (2) the difference in app behaviors. An example of the latter would be the inability to map Etsy's events \textsf{b1-1} and \textsf{b2-1}  to Wish in Figure~\ref{fig:example}.

\vspace{-2mm}
\subsection{Strategies Explored to Date}
\label{sec:bg:existing_work}

Four recent techniques \cite{behrang2018test, behrang2019atm, hu2018appflow, lin2019craftdroid} have  targeted UI test reuse in Android. 
The shared core concern of these techniques is to correctly map the GUI events from a source app to a target app. 
In the example from Figure~\ref{fig:example}, the source test \emph{sign-in} in Wish comprises the event sequence  \{\textsf{a1-1}, \textsf{a1-2}, \textsf{a1-3}\}. By mapping GUI events in this test from Wish to Etsy as  
\{\textsf{a1-1} $\rightarrow$ \textsf{b3-1}, \textsf{a1-2} $\rightarrow$ \textsf{b3-2}, \textsf{a1-3} $\rightarrow$ \textsf{b3-3}\},  
a new \emph{sign-in} test for the target app,  Etsy, is generated as \{\textsf{b3-1}, \textsf{b3-2}, \textsf{b3-3}\}.


Existing techniques can be classified into two main categories, based on 
how they map GUI events: 
\textbf{\appflow}~\cite{hu2018appflow} is ML-based, while \textbf{\craft}~\cite{lin2019craftdroid}, \textbf{\guitest}~\cite{behrang2018test}, and \textbf{ATM}~\cite{behrang2019atm} are similarity-based.
We have abstracted the two categories and their workflows by studying the similarities and differences of existing techniques.

\textbf{ML-based} techniques learn a classifier from a training dataset 
of different apps' GUI events  based on certain features, such as
text, element sizes, and image recognition results of graphical icons. 
The classifier is used to recognize  \emph{app-specific} GUI events and map them to  \emph{canonical} GUI events used in a test library, so that  {app-specific} tests can be generated by reusing the tests defined in the test library. 

\textbf{Similarity-based} techniques define their own algorithms to compute a \emph{similarity score} between pairs of GUI events in a source app and a target app based on the information extracted from the two apps, such as text and  element attributes. 
The similarity score is used to determine whether there is a match between each GUI event in the source app and the target app 
based on a customizable \emph{similarity threshold}. 
For example, \textsf{a1-1} in Wish (left) from Figure~\ref{fig:example} is likely to have a higher similarity score with \textsf{b3-1} than with other GUI events in Etsy (right). In that case, \textsf{a1-1} in Wish will be mapped to \textsf{b3-1} in Etsy.
Another important component in similarity-based techniques is the \emph{exploring strategy}, which determines the order of computing the similarity score between the GUI events in the source and   target apps. 
The target app's events that are explored earlier usually have a higher chance of being mapped.

\subsection{Existing Evaluation Metrics}
\label{sec:bg:existing_metrics}

To evaluate their test-reuse strategies, existing techniques have  focused on the \emph{accuracy} 
of the \emph{GUI event mapping}.
This section overviews the metrics they applied, which guided us in defining the expanded set of \framework's metrics 
(see Section~\ref{sec:design:metrics:accuracy}).
Note that the detailed definitions of existing metrics were not provided in the  publications~\cite{hu2018appflow,behrang2018test,behrang2019atm,lin2019craftdroid}; 
we had to separately contact the authors of each technique to obtain the details introduced below.

\textbf{\appflow}~\cite{hu2018appflow} is an ML-based technique that maps app-specific events to canonical events using a classifier as discussed earlier. 
\appflow's classifier is evaluated with the standard \emph{accuracy} metric~\cite{accuracy_precision_recall}, indicating the percentage of the correctly-classified GUI events  among all the GUI events being classified. 
Correctly-classified GUI events include two cases: (1) the app-specific events that are mapped to the correct canonical events (true positive); and (2) the app-specific events that are not mapped to any canonical events and such canonical events do not exist (true negative).

\textbf{\craft}~\cite{lin2019craftdroid} is a similarity-based technique. After the transfer of events from a source app to  events in a target app, {\craft}'s authors manually identify three cases: 
(1) \emph{true positive} (TP) occurs when the transferred event is the same as the one obtained during a manual transfer; 
(2) \emph{false positive} (FP) occurs when the transferred event is different; 
 and (3) \emph{false negative} (FN) occurs when \craft~fails
 to find a matching event, while the manual transfer succeeds. 
\emph{Precision} and \emph{recall} are then calculated based on the three cases. 
It is important to note that \craft's FP includes both the incorrectly transferred events and the newly added \emph{ancillary events} (if any), which is different from the FP case defined in other techniques.
We further illustrate this in Section~\ref{sec:design:metrics:accuracy}.

\textbf{ATM}~\cite{behrang2019atm} and \textbf{GTM}~\cite{behrang2018test} are also similarity-based techniques, and ATM is an enhancement of GTM by the same authors. 
Similar to \craft, the authors manually inspect the transferred results 
and identify four cases:
	(1) \emph{correctly matched} means the source event is mapped to the correct event in the target app (TP);
	(2) \emph{incorrectly matched} means the source event is mapped to the wrong event in the target app (FP);	
	(3) \emph{unmatched (!exist)} means the source event is not mapped to any  events and no such events exist in the target app (TN);
	(4) \emph{unmatched (exist)} means the source event is not mapped to any  events although the matching event exists in the target app (FN).
ATM and GTM do not calculate the precision or recall, but  present the raw percentages of each  of the four cases.

\section{\emph{F\lowercase{r}UIT\lowercase{e}R}'\lowercase{s} Principal Requirements}
\label{sec:design:requirements}

This section elaborates on the key limitations of  current test-reuse techniques and their evaluation processes. 
These limitations serve as the foundation of five  requirements we focused on in \framework's design (Section~\ref{sec:design}) and instantiation (Section~\ref{sec:foundation}). 


Prior to developing \framework, we investigated the existing techniques and their evaluations~\cite{hu2018appflow,lin2019craftdroid,behrang2018test,behrang2019atm} in depth.
Beyond consulting the available publications, we also studied the techniques' implementations and produced artifacts~\cite{appflowrepo,atm_repo,craftdroid_repo,gtm_repo}, and engaged their authors in, at times, extensive discussions to obtain missing details and resolve ambiguities.
In the end, we identified five limitations 
that are likely to hinder future advances in this emerging area.
We base \framework's principal requirements on these limitations. 

\textbf{Req$_1$ --- \emph{Metrics used by \framework to evaluate test-reuse techniques shall be standardized and reflect practical utility. --- }}
\noindent Existing techniques are evaluated with different, and differently applied, metrics (recall Section~\ref{sec:bg:existing_metrics}), which harms their side-by-side comparison. 
More importantly, all techniques to date have focused on whether  GUI events from a source app are correctly transferred to a target app, without considering whether the transferred \emph{tests} are actually meaningful and applicable in the context of the target app.
It is thus possible that all GUI events are mapped correctly, but the transferred test cannot be applied, e.g., due to missing ancillary events (recall Section~\ref{sec:bg:example_term}). None of the existing techniques are able to identify such scenarios; \framework must be able to do so.


\textbf{Req$_2$ --- \emph{\framework's workflow shall reduce the required manual effort and thus scale to larger numbers of apps and tests than possible with current test-reuse techniques. ---}}
\noindent Existing techniques' evaluation processes require significant manual effort to inspect every  transferred event in each test. 
For example, ATM~\cite{behrang2019atm} was evaluated on 4 app categories, where each category, in turn, consisted of  4 apps.
On average, each app had 10 tests to be transferred and each test had 5 events.
Within each app category, ATM transferred the tests of each app to the remaining 3 apps, resulting in 48 source-target app pairs in total.
For each app pair, ATM's authors had to manually inspect an average of 50 transferred events  (10 tests $\times$ 5 events), i.e., 2,400 events in total.
This is why they were forced to restrict their comparison with \guitest~\cite{behrang2018test} to a randomly selected half of  possible source-target app pairs. 
\framework must address this shortcoming by providing a more scalable evaluation workflow that requires markedly less manual effort. 

\textbf{Req$_3$ --- \emph{Evaluation results produced by \framework shall be reproducible. --- }} 
\noindent As discussed in Section~\ref{sec:bg:existing_metrics}, the current techniques' evaluation results depend on identifying the case to which each transferred event belongs (e.g., \emph{correctly matched}, \emph{false positive}, etc.).
Such ``ground-truth mappings'' are determined manually.
However, there are no standard guidelines for conducting   inspections, making the results potentially biased and unreproducible.
In Figure~\ref{fig:example}'s example, it is debatable whether  \{\textsf{a1-1} $\rightarrow$ \textsf{b3-1}\} is correct because \textsf{a1-1} only takes the user's email, while \textsf{b3-1} takes both the email and username.
ATM's authors also acknowledge the possibility of mistakes in the manual process~\cite{behrang2019atm}.
More importantly, any such mistakes are hard to locate or verify by other researchers, since the results of  manual inspection and the ground-truth mappings on which they are based,  are recorded in ad-hoc ways.
Thus, to facilitate future research in this area, the evaluation results produced by \framework must be reproducible, with a ground-truth representation that can be independently verified, reused, and modified. 

\textbf{Req$_4$ --- \emph{Test-reuse capabilities incorporated and evaluated by \framework shall be modularized. ---}}
\noindent Despite providing similar functionality, existing test-reuse techniques are designed as one-off solutions and evaluated as a whole. This makes it difficult to reuse or compare their relevant  components. In turn, it invites duplication of effort and introduces the risk of missed opportunities for advances by other researchers, and even by the techniques' own developers. 
To address this problem, \framework must modularize each test-reuse artifact it evaluates, allow its independent (re)use, and associate the obtained evaluation results with the appropriate artifacts. 

\textbf{Req$_5$ --- \emph{Benchmarks provided and applied by \framework shall be re\-usable. ---}}
\noindent Existing test-reuse techniques have been evaluated using different benchmark apps and tests, additionally hampering their comparison. 
In fact, only three subject apps were shared by two (\appflow~\cite{hu2018appflow} and \craft~\cite{lin2019craftdroid}) out of the four existing techniques  in their evaluations. 
The underlying reason is the different assumptions made by the techniques.
For instance, \guitest~ and ATM rely on the Espresso testing framework~\cite{espresso} that requires the apps' source code. 
As another example, AppFlow's tests are written in a special-purpose language based on Gherkin~\cite{Gherkin} and cannot be reused by techniques that capture tests  in other languages (e.g., Java, used by ATM and GTM). 
Thus, \framework must establish a set of uniform benchmarks with reusable apps and tests that can serve as the foundation for evaluating and comparing solutions in this  area.

\section{\emph{F\lowercase{r}UIT\lowercase{e}R}'\lowercase{s} Design}
\label{sec:design}

This section presents  \framework's design, with a focus on two  features that  address requirements Req$_1$, Req$_2$, Req$_3$, and partially Req$_4$: new evaluation metrics and an automated, modular workflow. We also introduce two novel test-reuse techniques to serve as baselines for bounding the existing techniques' evaluation results.

\subsection{\framework's Metrics}
\label{sec:design:metrics}

To address Req$_1$, \framework incorporates a pair of evaluation metrics: (1) {\emph{fidelity}}  focuses on how correctly the {GUI events are mapped}  from a source app to a target app;  (2) {\emph{utility}}  measures how useful the {transferred tests} are in practice.

\subsubsection{{Fidelity Metrics}} 
\label{sec:design:metrics:accuracy}

As explained in Section~\ref{sec:bg:existing_metrics}, fidelity  
of the mapping has been the main focus of existing techniques, but the previous metrics have been used inconsistently.\footnote{Existing publications in this area have referred to some of these as ``accuracy'' metrics. We use ``fidelity'' to avoid confusion with a specific  metric named ``accuracy'' defined previously in literature~\cite{accuracy_precision_recall} and used by one of the techniques we studied~\cite{hu2018appflow}.} 
To form a fair playground for comparing test-reuse techniques, we studied existing metrics by consulting available documentation and discussing with their  authors. 
We standardized this information into a \emph{comprehensive} set of fidelity metrics in \framework, as shown in Table~\ref{tbl:metric_mapping}.

\begin{table}[t!]
\centering
\caption{Fidelity metrics as used in App\-Flow~\cite{hu2018appflow}, \craft~\cite{lin2019craftdroid}, ATM~\cite{behrang2019atm}, GTM~\cite{behrang2018test}, and \framework.}
\label{tbl:metric_mapping}
\centering
\resizebox{\linewidth}{!}{
\begin{tabular}{|c|c|c|c|c|c|c|c|c|}
\hline
\textbf{\begin{tabular}[c]{@{}c@{}}~\\ ~\end{tabular}} & \begin{tabular}[c]{@{}c@{}}\textbf{True Pos.}\\ \textbf{(TP)}\end{tabular} & \multicolumn{2}{c|}{\begin{tabular}[c]{@{}c@{}}\textbf{False Pos.}\\ \textbf{(FP)}\end{tabular}} & \begin{tabular}[c]{@{}c@{}}\textbf{True Neg.}\\ \textbf{(TN)}\end{tabular} & \begin{tabular}[c]{@{}c@{}}\textbf{False Neg.}\\ \textbf{(FN)}\end{tabular} & \textbf{Accuracy}               & \textbf{Precision}                & \textbf{Recall}         \\  \hline
\hline
\textbf{AppFlow}                                            & \textcolor{cyan}{\emph{anon} }                                                    & \multicolumn{2}{c|}{\textcolor{cyan}{\emph{anon}}}                                                       & \textcolor{cyan}{\emph{anon} }                                                     & \textcolor{cyan}{\emph{anon} }                                                     & Accuracy          & \textcolor{brown}{\emph{dnc}}                & \textcolor{brown}{\emph{dnc}}             \\ \hline
\textbf{CraftDroid}                                         & TP                                                          & \multicolumn{2}{c|}{FP1}                                        & \textcolor{lightgray}{\emph{none}}                                                           & FN                                                          & \textcolor{lightgray}{\emph{none}}                & Precision          & Recall          \\ \hline
\textbf{\begin{tabular}[c]{@{}c@{}}ATM/\\ GTM\end{tabular}} & \begin{tabular}[c]{@{}c@{}}Correctly\\ Matched\end{tabular} & \multicolumn{2}{c|}{\begin{tabular}[c]{@{}c@{}}Incorrectly\\ Matched\end{tabular}} & \begin{tabular}[c]{@{}c@{}}Unmatched\\ (!exist)\end{tabular} & \begin{tabular}[c]{@{}c@{}}Unmatched\\ (exist)\end{tabular} & \textcolor{brown}{\emph{dnc}}               & \textcolor{brown}{\emph{dnc}}                & \textcolor{brown}{\emph{dnc}}             \\ \hline
\textbf{FrUITeR}                                            & Correct                                                     & \multicolumn{2}{c|}{Incorrect}                                                     & NonExist                                                     & Missed                                                      & Accuracy          & Precision          & Recall          \\ \hline
\end{tabular}
}
\end{table}

Table~\ref{tbl:metric_mapping} presents the fidelity metrics used across the different test-reuse techniques, and their relationship to the standard metrics as defined in literature~\cite{accuracy_precision_recall}.
Each row shows a mapping from the names for the metrics used by each technique to the typical fidelity metrics' names indicated in the header.
``\textcolor{cyan}{\emph{anon}}'' cells represent metrics that are not reported by a technique, but are used internally to calculate other metrics that are reported. ``\textcolor{brown}{\emph{dnc}}'' cells represent  metrics that are not calculated by a given technique, but can be determined based on other metrics used. Finally, ``\textcolor{lightgray}{\emph{none}}'' cells represent cases where a  metric is not used by a technique and cannot be calculated from the available information. 
\framework~ covers all seven metrics, changing several metrics' names  to better reflect their application to test reuse, as will be further discussed below.

Recall from Section~\ref{sec:bg:existing_metrics} that \craft's FP category covers two cases: FP1 corresponds to ``Incorrectly Matched'' events in ATM/GTM and ``Incorrect'' in \framework; FP2 corresponds to the ancillary events that are not considered by other techniques.
\framework also excludes the ancillary events from its Incorrect category because they can be benign or even needed (e.g., \textsf{b1-1} and \textsf{b2-1} from Figure~\ref{fig:example}), and do not reflect the fidelity of the {GUI event mapping}.
For instance, if ancillary events were considered to be False Positives, a large number of them would result in a low Precision  for the GUI event mapping. However, this would not be a meaningful measure since the ancillary events are not mapped from the source app.
Such events are thus not relevant to the mapping's \emph{fidelity}, but should be  considered by the \emph{utility} metrics, introduced next.

\subsubsection{{Utility Metrics}} 
\label{sec:design:metrics:utility}

\framework introduces two \emph{utility} metrics to indicate how useful a \emph{transferred test} is. 
This aspect is not considered in prior work, but 
is needed because a high-fidelity event mapping does not guarantee a successfully transferred test, or vice versa.
For instance, a target app's ground-truth test may contain \emph{ancillary events} not covered by  source events, making it impossible to generate a ``perfect'' test by event mapping alone.
On the flip side, a low-fidelity mapping may accidentally generate a ``perfect'' test. 
Thus, it is important to measure the utility 
with respect to the ground-truth test \emph{independently} of  event mapping's fidelity.

To this end, we first define an \emph{effort} metric, to
measure how close the transferred test is to the ground-truth test, by calculating the two tests' Levenshtein distance~\cite{levenshtein1966binary}.  
Levenshtein distance is widely used in NLP  to measure the  steps needed to transform one string into another. 
In our case, each step is defined as the \emph{insertion}, \emph{deletion}, or \emph{substitution} of an event in the transferred test.


Secondly, we  define a \emph{reduction} metric, to assess the manual effort saved by the generation of the transferred test,  compared to writing the ground-truth test from scratch: \\
\indent\indent\indent \textbf{Reduction} = (\#gtEvents -- effort) $\div$ \#gtEvents\\
The value of \emph{reduction} may be negative, if transforming the transferred test into the corresponding ground-truth test 
takes more steps than constructing such ground-truth test from scratch. 

Note that each usage-based test targets a scenario with a specific flow of interest; multiple flows would result in multiple tests (e.g., sign-in from ``homepage'' vs.  from ``settings''). 
For each particular flow, it is possible for the  ground-truth test to contain different ``acceptable'' ancillary events based on one's interest, which would result in multiple ``acceptable'' ground-truth tests. 
In \framework's current benchmark (see Section \ref{sec:foundation:benchmark}), we manually constructed one ground-truth test for each flow with the minimal amount of ancillary events to match prior work.
However,  \framework's ground-truth tests can be  modified or extended to obtain their corresponding utility results.
For instance, researchers can specify multiple ``acceptable'' ground-truth tests for a given flow, and measure the transferred test's utility with respect to each ground-truth test.

We acknowledge that the utility aspect (i.e., how useful a transferred test is) can be subjective depending on one's goal.
Alternative utility metrics (e.g., bug-identification power, executability, code coverage) can be added to \framework's customizable workflow (see Section \ref{sec:design:workflow}).
\framework's current utility metrics specifically center around \emph{effort} because they are applied to tests transferred by techniques whose end-goal is to reduce the effort of writing tests manually.
Refining utility's definition with extended  metrics is worthy of further study, but outside our scope. 
Our goal was to show that utility is important and measurable, to 
motivate further exploration of such important aspect that has been missed by prior work.

\subsection{F\lowercase{r}UIT\lowercase{e}R'\lowercase{s} Workflow}
\label{sec:design:workflow}

To address {Req$_2$}, {Req$_3$}, and partially {Req$_4$} from Section~\ref{sec:design:requirements}, we designed an \emph{automated} evaluation workflow with \emph{customizable} components,  shown in Figure~\ref{fig:workflow}. The goal of \framework's workflow is to generate \emph{reproducible} evaluation results for a test-reuse technique's core functionality. 
The workflow's automation is enabled by two key aspects: (1) the \emph{uniform representation} of the inputs and artifacts needed in the evaluation process, and (2)~a set of \emph{customizable components} that output the evaluation results of interest automatically.

\subsubsection{{Uniform Representation of Inputs}}
\label{sec:design:workflow:representation}

\begin{figure}[t]
	\centering
		\includegraphics[width=0.45\textwidth]{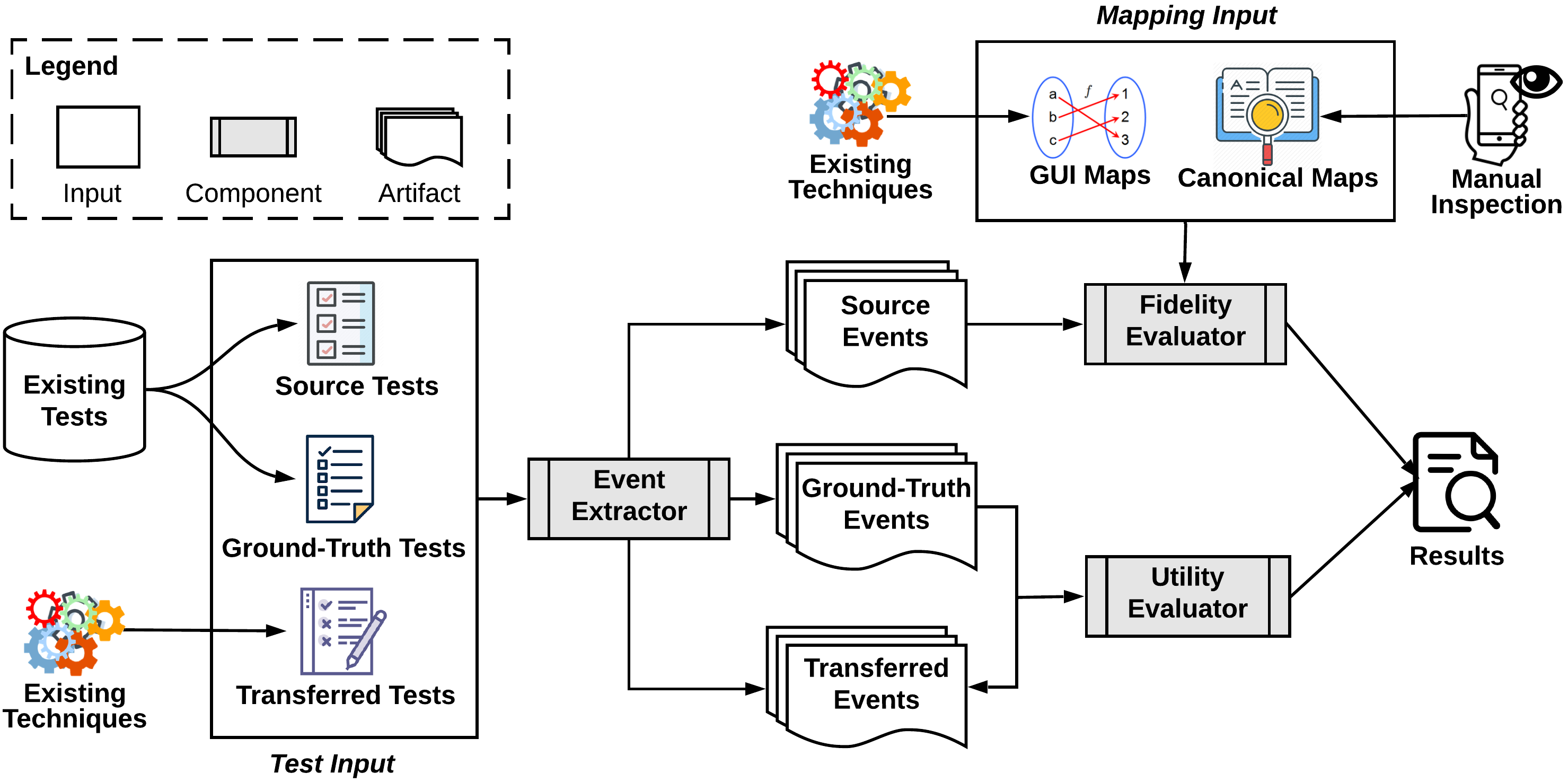}
	\caption{Overview of \framework's automated workflow.}
	\label{fig:workflow}
\end{figure}

As Figure~\ref{fig:workflow} shows, \framework takes two types of input: Test Input (bottom-left) and Mapping Input (top-right). The two  are a combination of inputs taken and artifacts produced by existing test-reuse techniques, as well as three new inputs  introduced in \framework~ to automate the evaluation process: Ground-Truth Tests, GUI Maps and Canonical Maps.

\textbf{Test Input} contains  source tests, ground-truth tests, and transferred tests as defined in Section~\ref{sec:bg:example_term}.
The tests may be captured in various forms by a test-reuse technique, and cannot be analyzed in a standard way. 
For instance, all tests in ATM~\cite{behrang2019atm} and GTM~\cite{behrang2018test} are represented as Espresso tests~\cite{espresso} in Java, while \craft~\cite{lin2019craftdroid}'s source tests are written in Python using Appium~\cite{appium} and its transferred tests are represented in JSON~\cite{json}. 
In order to enable their automated evaluation, the heterogeneous tests thus need to be standardized. 
\framework's Event Extractor converts the Test Input into a uniform representation of source events, ground-truth events, and transferred events 
as detailed in Section~\ref{sec:design:workflow:comp}.

\textbf{Mapping Input} consists of the GUI Map and the Canonical Map, for automatically  evaluating a test-reuse technique's \emph{fidelity}. The two maps are newly introduced by \framework and captured using a standardized representation. 
The \emph{GUI Map} contains the \emph{GUI event mapping} from a source app to a target app generated by a given test-reuse technique, and is used to compute the {fidelity} metrics introduced in Section~\ref{sec:design:metrics:accuracy}.
Prior work does not provide GUI Maps, but only the final Transferred Tests. The events in these tests cannot be used to calculate fidelity by comparing with source events directly, because the transferred events may include \emph{ancillary} and \emph{null} events.
We further illustrate how we extract the GUI Maps from existing techniques and evaluate their fidelity automatically with \framework in Section~\ref{sec:foundation}.
On the other hand, the \emph{Canonical Map} contains the mapping from  app-specific events to  canonical events. This map is manually constructed and is used as the ground-truth mapping for \framework's Fidelity Evaluator component discussed below.
Note that \appflow~\cite{hu2018appflow} can generate a Canonical Map automatically using ML techniques. However, \appflow's certain mapping results can be wrong, and thus cannot be used as the ground truth.

\subsubsection{{Customizable Components}}
\label{sec:design:workflow:comp}

\framework~ introduces three customizable components, shown as shaded boxes in  Figure~\ref{fig:workflow}: Event Extractor, Fidelity Evaluator, and Utility Evaluator. 

\textbf{Event Extractor} leverages program analysis to extract the GUI event sequence from 
the  tests' code.
The  sequence is represented as each event's \textsf{ID} or \textsf{XPath}, depending on which of the two is used in the test. 
\textsf{ID} and \textsf{XPath} are widely used to locate specific GUI elements in tests in various domains, including Android apps~\cite{android_find_elements} and web apps~\cite{webidxpath}.  
For simplicity, we will use ``\textsf{ID}'' to refer to either the \textsf{ID} or \textsf{XPath} of a specific event in the rest of the paper.

To extract the event sequence, Event Extractor  analyzes the {Test Input} to  locate the program point of each event based on its corresponding API, e.g., \textsf{click}~\cite{click} or \textsf{sendKeys}~\cite{sendkeys} for tests written with   Appium~\cite{appium}.
Once it identifies the location, Event Extractor determines the event's {caller}, i.e., the GUI element where the event is triggered, and performs a def-use analysis~\cite{allen1976program} to trace back the definition of the caller's \textsf{ID}. 
This definition is specified in a given API of the testing framework, such as  \textsf{findElementById()} in Appium~\cite{find_element_id}.
In that case, the def-use analysis is used to pinpoint the \textsf{findElementById()} call that corresponds to the event's caller so that \textsf{ID}'s value can be determined.
The input value associated with the event (if any) is determined by def-use analysis in the same manner. 
In the end, the converted Source Events, Ground-Truth Events, and Transferred Events are represented in a uniform way with \textsf{ID}s regardless of what testing framework is used. 

Note that if the Test Input is written in different  languages or testing frameworks, multiple Event Extractor instances need to be implemented. 
However, this is a one-time effort, and subsequent  work can reuse existing Event Extractors when applied on the tests written in the same language and testing framework.
Moreover, the Event Extractor is easily customizable to process tests written with different  frameworks by replacing the relevant APIs' signatures.
For instance, when identifying an event caller's \textsf{ID}, 
the relevant API is \textsf{findElementById()} if using Appium \cite{appium} to test mobile apps, or \textsf{findElement()} if using Selenium \cite{selenium,webelement} to test web apps.
By simply replacing the relevant API signature, the Event Extractor will be able to process tests written in both Appium and Selenium.

\setlength{\textfloatsep}{0.15cm}
\setlength{\floatsep}{0.15cm}
\begin{algorithm}[t!]
\begin{spacing}{0.85}
\small
\DontPrintSemicolon 
\KwIn{EventList $srcEvents$, GUIMap $guiMap$, \qquad \quad
 CanonicalMap $srcCanMap$, $tgtCanMap$}
\KwOut{Sets $correct$, $incorrect$, $missed$, $nonExist$}

$correct = incorrect = missed = nonExist = \emptyset$


\For{$ i = 1$ to $srcEvents.size$}{
		$src \gets srcEvents.\textsc{GET}(i)$\\
		$trans \gets guiMap.\textsc{GetMapped}(src)$\\
		$srcCan \gets srcCanMap.\textsc{GetCanonical}(src)$\\
		$transCan \gets tgtCanMap.\textsc{GetCanonical}(trans)$\\
		\If{$trans$ != $null$}{ 
			\If{$transCan$ == $srcCan$} {
				$correct.\textsc{Put}(src)$
			}
			\Else{
				$incorrect.\textsc{Put}(src)$
			}		
		}
		\Else{ 
			\If{$tgtCanMap.\textsc{contains}(srcCan)$}{
				$missed.\textsc{Put}(src)$
			}
			\Else{
				$nonExist.\textsc{Put}(src)$
			}
		}		
				
}

\Return{$correct$, $incorrect$, $missed$, $nonExist$}
\caption{\sc Fidelity Evaluator}
\label{alg:accuracy}
\end{spacing}
\end{algorithm}

{\textbf{Fidelity Evaluator}} takes the {Source Events} produced by Event Extractor and Mapping Input, and automatically outputs the sets of (1) \emph{correct}, (2) \emph{incorrect}, (3) \emph{missed}, and (4) \emph{nonExist} cases for calculating \framework's seven fidelity metrics (recall Table~\ref{tbl:metric_mapping}).

Algorithm~\ref{alg:accuracy} describes {Fidelity Evaluator} in detail.
The algorithm iterates through each {source event} to determine to which of the four cases it should be assigned (Lines 2-16).
To do so, it first gets the current {source event} ($src$), and the transferred event mapped from it ($trans$) based on the GUI Map (Lines 3-4).
It then converts the app-specific events $src$ and $trans$ into their corresponding canonical events $srcCan$ and $transCan$, using their respective Canonical Maps, so that the events are comparable (Lines 5-6).
Finally, to determine which of the four cases $src$ falls into, the algorithm first checks whether $trans$ is a \emph{null} {event}. If 
 not, $transCan$ will be compared against $srcCan$ to determine whether the {transferred event} refers to the same canonical event as the {source event}, and $src$ will be added to either the \emph{correct}  or \emph{incorrect} set accordingly (Lines 7-11).
If $trans$ is \emph{null}, the {source event} has not been mapped to any events in the target app. The algorithm then iterates through  the {Canonical Map} of the target app ($tgtCanMap$) to determine whether the matching event $srcCan$ exists in the target app, and $src$ will be added to either the \emph{missed} set or \emph{nonExist} set accordingly  (Lines 12-16).


{\textbf{Utility Evaluator}} automatically analyzes the {Ground-Truth Events} and  {Transferred Events} produced by Event Extractor. It uses this information to compute the two utility metrics---\emph{effort} and \emph{reduction}---based on their definitions described in Section~\ref{sec:design:metrics:utility}. 



\subsubsection{Relationship to \framework's  Principal Requirements}
\label{sec:design:workflow:benefits}

\framework's \linebreak workflow yields three key  benefits that target {Req$_2$}, {Req$_3$}, and {Req$_4$}. 

First, the only manual effort required by \framework is  to construct the {Canonical Maps} by relating app-specific events to canonical events.
This  is a one-time effort \emph{per app}, and each event only needs to be labeled once regardless of how many times it appears in a test ({Req$_2$}). 
By contrast, in previous work~\cite{lin2019craftdroid,behrang2018test,behrang2019atm}, each app-specific event needs to be manually labeled every time it appears in a test, pos\-sibly resulting in thousands of manual inspections. 

Second, \framework~establishes ground truths with uniform representations: Canonical Maps are the ground truth for assessing \emph{fidelity}, while Ground-Truth Events help to assess \emph{utility}. This renders the evaluation results yielded by \framework reproducible ({Req$_3$}).
For instance, any  mistakes or subjective judgments made in the current techniques' manual evaluation processes can be easily located by inspecting the Canonical Maps, and independently reproduced.
Further, \framework's Canonical Maps are reusable, modifiable, and extensible for subsequent studies, helping to avoid duplicated work. 

Third, \framework's workflow consists of customizable modules that isolate the evaluation to a relevant  component of a test-reuse technique ({Req$_4$}).
For instance,  Fidelity Evaluator only assesses the performance of  GUI event mapping, instead of evaluating a technique as a whole.
Moreover, both Fidelity Evaluator  and Utility Evaluator can be customized, reused, or extended to automatically evaluate other metrics of interest based on  the standardized inputs and artifacts that \framework~ defines, directly fostering future research.

\subsection{\framework's Baseline Techniques}
\label{sec:design:metrics:baseline}

To better understand the performance of a test-reuse technique, we developed two baseline techniques---\emph{Na\"ive} and \emph{Perfect}---that establish the lower- and upper- bounds achievable by the fidelity and utility metrics in a given scenario.

\begin{algorithm}[b!]
\begin{spacing}{0.85}
\small
\DontPrintSemicolon 
\KwIn{EventList $srcEvents$, AppInfo $tgtAppInfo$}
\KwOut{EventList $transEvents$}

$transEvents \gets \emptyset$\\
$currentAct \gets tgtAppInfo.\textsc{getMainActivity}()$\\

\ForEach{$src \in srcEvents$}{
	$isMapped \gets FALSE$\\
	$events \gets tgtAppMap.\textsc{getAllevents}(currentAct)$\\
	$events.\textsc{randomizeOrder()}$\\
	\ForEach{$event \in events$}{
		\If{$event.\textsc{action} == src.\textsc{action}$}{
			$similarity \gets \textsc{getRandomSimilarity}(0$, $ 1)$\\
			\If{$similarity > Threshold$}{				
				$transEvents.\textsc{add}(event)$\\
				$currentAct \gets event.\textsc{nextActivity}()$\\
				$isMapped \gets TRUE$\\
				\textbf{break}
				}	
			}		
	}
	\If{$\neg isMapped$}{
		$transEvents.\textsc{add}(null)$
	}	
}
\Return{$transEvents$}
\caption{\sc Na\"ive Baseline Technique}
\label{alg:random}
\end{spacing}
\end{algorithm}

\subsubsection{{Na\"ive Baseline}} The Na\"ive baseline uses a random strategy to select the events in a target app to which each source event should be mapped. This sets the practical lower-bound of \emph{fidelity}.
As Algorithm~\ref{alg:random} shows, Na\"ive initially explores the target app from the main Activity \cite{AndroidActivity}~(Line 2). 
For each source event, it  obtains all  the events  at the current Activity ($events$) in a random order~(Lines 5-6), and then tries to find a match between the current source event $src$ and each event in $events$ (Lines 7-14).
When mapping $src$ to $event$, Na\"ive first checks if the associated actions of the two events are the same, and only computes the similarity score when they are.
The similarity score is computed by selecting a random value between 0~and~1~(Line~9), which are the lower and upper bounds used in existing work.
If the similarity score of $src$ and $event$ is above a certain threshold, 
$event$ is added to the list maintained in $transEvents$~(Line 11). At that point, Na\"ive  continues to explore the target app from the Activity reached by the transferred $event$~(Line~12), and  marks the current source event $src$ as mapped~(Line 13).
In the end, if the source event is not mapped, it will be marked as a \emph{null event} and added to $transEvents$ (Line~15-16). Null events correspond to either the True Negative or False Negative categories in Table~\ref{tbl:metric_mapping}. 

\subsubsection{{Perfect Baseline}} 

The {Perfect} baseline transfers the source events based on the ground-truth mapping (recall Section~\ref{sec:design:workflow}), assuming all source events are correctly mapped to the target app. 
Perfect baseline thus represents a ``perfect'' GUI event mapping and achieves 100\% fidelity by definition.
Specifically, we are interested in the \emph{utility} achieved by the Perfect baseline since it represents the upper-bound of the transferred tests' practical usefulness, which is not considered previously. 
This can help us identify the room for improvement and guide future research in test-reuse techniques.

\section{F\lowercase{r}UIT\lowercase{e}R'\lowercase{s}  Instantiation}
\label{sec:foundation}

This section describes how we instantiate \framework to automatically evaluate the relevant modules of existing techniques alongside \framework's baseline techniques, in partial satisfaction of {Req$_4$}. 
The evaluation is  based on \framework's reusable benchmark that addresses {Req$_5$}.
To this end, we needed to provide  information to enable \framework's workflow (recall Figure~\ref{fig:workflow}): Source Tests that are supplied as inputs to a given test-reuse technique; Transferred Tests and GUI Maps, which are produced as outputs of a given test-reuse technique; and the manually constructed ground truths, namely, Canonical Maps and Ground-Truth Tests.
However, existing test-reuse techniques were not designed with \framework's modular workflow in mind, and thus do not provide such information directly.

Section~\ref{sec:foundation:comparison} explains how we mitigated the above challenge in order to extract the relevant components from existing techniques and generate the information needed by \framework.
Note that this step will not be necessary for future techniques if they follow \framework's modularized design.
Section~\ref{sec:foundation:benchmark} presents  \framework's reusable benchmark for the uniform evaluation of  test-reuse techniques, which contains the Source Tests, Ground-Truth Tests, and Canonical Maps used in \framework's automated workflow. Finally, Section~\ref{sec:impl} provides the details of \framework's implementation and generated datasets.

\subsection{Modularizing Existing Techniques}
\label{sec:foundation:comparison}

To lay the foundation for addressing Req$_4$, we  modularized \framework's design. In turn, this isolated the evaluation of GUI event mapping's {fidelity} and the transferred tests' {utility}, as discussed in Section~\ref{sec:design:workflow}. However, the existing techniques are implemented and evaluated as fully integrated, one-off solutions that do not provide the artifacts needed by \framework~ to generate the modularized evaluation results.
Because of this, we had to extract the specific functionality  from existing techniques' implementations that performs the GUI event mapping (recall Section~\ref{sec:bg:existing_work}). 
Once the GUI Maps are available, we can generate the Transferred Tests used in \framework's Utility Evaluator. 
Note that the step of extracting  GUI Mapper components 
is not needed for future test-reuse techniques if they follow \framework's modularized design. 
For example, we directly applied \framework~ on the two  baseline techniques we developed, 
with no extra effort.

Extracting the GUI Mapper components from the existing techniques was  challenging since we had to understand each technique's design  and implementation  in detail, and to modify its source code.
To this end, in addition to the available publications, we studied in depth the   existing approaches' implementations~\cite{appflowrepo, atm_repo, gtm_repo, craftdroid_repo} and communicated with their authors extensively.
We describe the challenges we faced during this process and the specific component-extraction strategies we applied to each existing solution. 

\subsubsection{{Extracting \appflow's GUI Mapper}}
{\appflow}~\cite{hu2018appflow} is an ML-based technique whose key component trains a  classifier that maps app-specific events to  canonical events, but does not map the  events from a source app to a target app. 
To compare \appflow~ with  similarity-based techniques, we leverage its Canonical Maps to transfer the source events to the target app by (1) mapping each source event to the corresponding canonical event based on the source app's Canonical Map and (2) mapping this canonical event back to the app-specific event in the target app 
based on the target app's Canonical Map. 
\appflow's implementation does not output its Canonical Maps, so we had to locate and modify the relevant component to do so. 
Moreover, \appflow~ does not store its trained classifier, so we had to configure its ML model and re-train it.
During this process, we communicated with \appflow's authors closely to understand its code, to obtain proper configuration files and training data, and to ensure the correctness of our re-implementation.

\subsubsection{{Extracting ATM's and GTM's GUI Mappers}}

As discus\-sed earlier, {ATM}~\cite{behrang2019atm} was developed as an enhancement to {GTM}~\cite{behrang2018test} and was shown to outperform it~\cite{behrang2019atm}.
However, the authors of these two techniques compared them only on half of the source-target app pairs used in ATM's publication~\cite{behrang2019atm} due to the large manual effort required. 
Since \framework~ largely automates the comparison process, we decided to extract the GUI Mapper components from both techniques to enable their comparison at a large scale. 

An obstacle we had to overcome was that ATM and GTM both require the app's source code due to the use of the Espresso \cite{espresso}. Thus, they cannot be compared as-is with techniques evaluated on closed-sourced apps, which
would have limited our choice of benchmark apps. We discussed this with ATM's and GTM's authors and learned that the only step that requires  source code for  both techniques' GUI Mappers   is computing the textual similarity score of image GUI elements (e.g., \textsf{ImageButton}). In that case, the text of the image's filename is retrieved from the app's  code and analyzed to compute the  similarity score. However, the main author confirmed that, in her experience, this feature is  rarely needed in practice.
We thus decided to extract ATM's and GTM's GUI Mapper components
as  stand-alone Java programs that do not require Espresso, omitting the filename-retrieval feature.
We subsequently confirmed with the two techniques' authors the correctness of our implementation.

\subsubsection{{Extracting \craft's GUI Mapper}}

{\craft}'s~\cite{lin2019craftdroid}
implementation is only partially available. Its authors informed us that two of \craft's modules---Test Augmentation and Model Extraction---were not releasable when we requested them, due to ongoing modifications.
The authors confirmed our observation that \craft's  GUI mapping functionality depends on the outputs of the two missing modules, and advised us that the best strategy would be for us to reimplement them based on \craft's lone publication~\cite{lin2019craftdroid}. 
However, the publication in question is missing implementation details that would introduce bias: 
 we would have no guarantee that the versions of the two components we produce are the same as those used in \craft. 
Instead, we decided to rely on \craft's published Transferred Tests~\cite{craftdroid_repo} in our evaluation.

To obtain \craft's GUI Maps, we inspected its published artifacts~\cite{craftdroid_repo} and found that only certain events in the Transferred Tests have associated similarity scores, while other events are labeled as ``empty''. 
Further investigation showed that each event in the Transferred Tests belongs to one of  three cases: (1)  events with available similarity scores are successfully mapped from the source events; (2)  ``empty'' events are mapped from the source events but no match is found by \craft~(i.e., null events); (3)~the remaining events are not mapped from the source events but are added by \craft~(i.e., ancillary events).
We  excluded the ancillary events so that the resulting transferred events have a 1-to-1 mapping from the source events, giving us \craft's GUI Maps.

\subsection{\framework's Benchmark}
\label{sec:foundation:benchmark}

As discussed above in the motivation for {Req$_5$}, existing test-reuse techniques are evaluated on different apps and tests, which  hinders their comparability.
To address this, we established a reusable benchmark   with the same apps and tests to serve as a shared measuring stick in this emerging domain.
This section discusses our strategy for including existing apps and tests in the benchmark, and for generating the required ground truth.

\subsubsection{{Benchmark Apps and Tests}}
To maximize the results from existing work that we can attempt to reproduce, we first included the intersection of the subject apps used by existing work. This yielded 3 shopping apps: Geek, Wish, and Etsy.
We further randomly selected 7 additional  shopping apps and 10 news apps used by \appflow~\cite{hu2018appflow}. This gave us 20 benchmark apps in total, as described in Table~\ref{tbl:subject_apps}.
Our rationale behind this choice of apps was two-fold: (1) \appflow's authors manually inspected all app categories on Google Play and identified shopping and news  as  categories with common functionalities suitable for test reuse; (2)~\appflow~ was evaluated on the largest number of subject apps among the existing techniques. By comparison, ATM~\cite{behrang2019atm} used 16 open-source apps that are not as popular as those used in \appflow.

\begin{table}[b!]
\centering

\caption{Summary information of benchmark apps.}
\label{tbl:subject_apps}
\centering
\resizebox{.9\linewidth}{!}{
\begin{tabular}{|l|c|c|c|c|c|}
\hline
\multirow{11}{*}{\textbf{Shopping}}                   & \textbf{App ID} & \textbf{App Name} & \textbf{\#Downloads} & \textbf{\#Tests} & \textbf{\#Events} \\ \cline{2-6}
                                                      & S1              & AliExpress        & 100M                 & 15               & 76                \\ \cline{2-6}
                                                      & S2              & Ebay              & 100M                 & 13               & 48                \\ \cline{2-6}
                                                      & S3              & Etsy              & 10M                  & 13               & 55                \\ \cline{2-6}
                                                      & S4              & 5miles            & 5M                   & 12               & 78                \\ \cline{2-6}
                                                      & S5              & Geek              & 10M                  & 13               & 85                \\ \cline{2-6}
                                                      & S6              & Google Shopping   & 1M                   & 15               & 72                \\ \cline{2-6}
                                                      & S7              & Groupon           & 50M                  & 14               & 66                \\ \cline{2-6}
                                                      & S8              & Home              & 10M                  & 14               & 98                \\ \cline{2-6}
                                                      & S9              & 6PM               & 500K                 & 14               & 63                \\ \cline{2-6}
                                                      & S10             & Wish              & 100M                 & 14               & 85                \\ \hline
\multicolumn{1}{|c|}{\multirow{10}{*}{\textbf{News}}} & N1              & The Guardian      & 5M                   & 13               & 76                \\ \cline{2-6}
\multicolumn{1}{|c|}{}                                & N2              & ABC News               & 5M                   & 9                & 31                \\ \cline{2-6}
\multicolumn{1}{|c|}{}                                & N3              & USA Today         & 5M                   & 11               & 28                \\ \cline{2-6}
\multicolumn{1}{|c|}{}                                & N4              & News Republic     & 50M                  & 10               & 40                \\ \cline{2-6}
\multicolumn{1}{|c|}{}                                & N5              & BuzzFeed          & 5M                   & 11               & 50                \\ \cline{2-6}
\multicolumn{1}{|c|}{}                                & N6              & Fox News          & 10M                  & 11               & 28                \\ \cline{2-6}
\multicolumn{1}{|c|}{}                                & N7              & SmartNews         & 10M                  & 9                & 20                \\ \cline{2-6}
\multicolumn{1}{|c|}{}                                & N8              & BBC News          & 10M                  & 9                & 22                \\ \cline{2-6}
\multicolumn{1}{|c|}{}                                & N9              & Reuters           & 1M                   & 10               & 37                \\ \cline{2-6}
\multicolumn{1}{|c|}{}                                & N10             & CNN               & 10M                  & 9                & 24                \\ \hline
\end{tabular}
}
\end{table}

To construct the benchmark tests, we further followed the test cases defined in \appflow, with a similar rationale:
(1) \appflow's authors conducted an extensive study to manually identify  tests that are shared in shopping  and news apps; (2) \appflow~ defines a larger number of tests compared to other work. For example, \craft~\cite{lin2019craftdroid} only has 2 tests defined in each app category.
We  excluded those tests that require mocking external dependencies (e.g., a payment service). This resulted in  15 tests in the shopping category and 14 tests in the news category, shown in Table~\ref{tbl:test}. 
Note that we cannot reuse \appflow's tests directly because they are written in a special-purpose language defined by \appflow~ for an entire app category rather than  a specific app.
Instead,
we relied on multiple undergraduate and graduate students with Android experience to write  the applicable tests for each of the 20 subject apps using Appium~\cite{appium}. 
Some benchmark apps did not have each functionality described in Table~\ref{tbl:test}, ultimately resulting in a total of 239 tests involving 1,082 events across the 20 apps (the two right-most columns of Table~\ref{tbl:subject_apps}), requiring 3,920 SLOC of Java code.

These benchmark tests currently do not contain oracle events because only ATM \cite{behrang2019atm} and \craft~\cite{lin2019craftdroid} can transfer oracles in principle.
However, due to the limited availability of \craft's source code as mentioned earlier, we would not be able to obtain \craft's results using our benchmark, making a comparison  across different techniques impossible.
As additional test-reuse techniques are developed with the ability to transfer oracles, \framework's benchmark tests can be extended to include oracle events to obtain their results as well.
Note that as long as future techniques follow \framework's modularized design to provide the needed input (e.g., GUI Maps of the oracle event mapping), \framework will be able to automatically generate the results of oracle events.
A detailed tutorial is provided on \framework's website \cite{fruiterrepo}.

\begin{table}
\centering
\caption{Benchmark test cases in shopping (TS) and news (TN) categories.}
\label{tbl:test}
\centering
\resizebox{\linewidth}{!}{
\begin{tabular}{|c|c|c|}
\hline
\textbf{Test ID} & \textbf{Test Case Name} & \textbf{Tested Functionalities}                       \\ \hline
TS1/TN1          & Sign In                 & provide username and password to sign in              \\ \hline
TS2/TN2          & Sign Up                 & provide required information to sign up               \\ \hline
TS3/TN3          & Search                  & use search bar to search a product/news               \\ \hline
TS4/TN4          & Detail                  & find and open details of the first search result item \\ \hline
TS5/TN5          & Category                & find first category and open browsing page for it     \\ \hline
TS6/TN6          & About                   & find and open about information of the app            \\ \hline
TS7/TN7          & Account                 & find and open account management page                 \\ \hline
TS8/TN8          & Help                    & find and open help page of the app                    \\ \hline
TS9/TN9          & Menu                    & find and open primary app menu                        \\ \hline
TS10/TN10        & Contact                 & find and open contact page of the app                 \\ \hline
TS11/TN11        & Terms                   & find and open legal information of the app            \\ \hline
TS12             & Add Cart                & add the first search result item to cart              \\ \hline
TS13             & Remove Cart             & open cart and remove the first item from cart         \\ \hline
TS14             & Address                 & add a new address to the account                      \\ \hline
TS15             & Filter                  & filter/sort search results                            \\ \hline
TN12             & Add Bookmark            & add first search result item to the bookmark          \\ \hline
TN13             & Remove Bookmark         & open the bookmark and remove first item from it \\ \hline
TN14             & Textsize                & change text size                                       \\ \hline
\end{tabular}
}
\end{table}

\subsubsection{{Benchmark Ground Truth}}
As described in Section~\ref{sec:design:workflow}, 
 we define Canonical Maps to represent the {ground truth} for the \emph{fidelity} of the GUI event mapping, 
and Ground-Truth Events to represent the ground truth for the \emph{utility} of the transferred tests.

In our benchmark, we define 72 canonical events for the shopping apps and 55 for the news apps.
Our canonical events are extended from \appflow, aiming to reflect a finer-grained classification of GUI events.
For instance, event ``\textsf{password}'' in the \emph{sign-in} test (TS1/TN1 in Table~{\ref{tbl:test}}),
and events ``\textsf{password}'' and ``\textsf{confirm password}'' in the \emph{sign-up} test (TS2/TN2 in Table~{\ref{tbl:test}}), 
 are all represented as the same canonical event ``\textsf{Password}'' in \appflow.
However, it is debatable whether that is appropriate. For example, mapping  ``\textsf{password}'' in \emph{sign-up} to  ``\textsf{password}'' in \emph{sign-in} may lead to non-executable tests.
To remove ambiguity, we capture such events separately. 

Based on the canonical events, we construct 20 Canonical Maps, one per subject app. We do so by manually relating to the canonical events a total of 561 subject apps' GUI events that appear in one or more of the 239 tests. As discussed in Section~\ref{sec:design:workflow}, this is the only manual step required by \framework and  is a one-time effort:
the Canonical Maps can be  reused when relying on the same subject apps. 
As a point of comparison, recall from Section~{\ref{sec:design:requirements}} that evaluating 48 app pairs in ATM~{\cite{behrang2019atm}} required manually inspecting 2,400 events. By contrast, our one-time inspection of the 561 events enabled  the use  of 200 app pairs (2 categories $\times$ 10$\times$10 apps, i.e., including an app's test transfer to itself) by every technique \framework evaluated. 

The Ground-Truth Events in our benchmark are extracted from the 239  tests 
by \framework's Event Extractor (recall Figure~\ref{fig:workflow}).

\subsection{F\lowercase{r}UIT\lowercase{e}R'\lowercase{s} Implementation Artifacts}
\label{sec:impl}

\framework's artifacts are publicly available~\cite{fruiterrepo}: its  source code; final datasets; GUI Mappers  extracted from existing work;  implementations of  baseline techniques, their  GUI Maps, and Transferred Tests;  benchmark apps and tests; and   manually constructed benchmark ground truths. 
We highlight the key details of these artifacts below.

\subsubsection{{Source Code}}
\framework's Event Extractor (recall Figure~\ref{fig:workflow}) is implemented in Java using Soot \cite{soot} (235 SLOC). \framework's Fidelity Evaluator and Utility Evaluator are implemented in Python (1,045 SLOC).
\framework's baseline techniques Na\"ive and Perfect (recall Section~\ref{sec:design:metrics:baseline}) are likewise implemented in Python (112 SLOC).
The  GUI Mapper components extracted from existing techniques (recall Section~\ref{sec:foundation:comparison}) are implemented in their original programing languages: \appflow~ in Python (1,084 SLOC); GTM in Java (1,409 SLOC); and ATM in Java (1,314 SLOC). 
The functionality that processes their outputs and generates the uniform representation of GUI Maps and Transferred Tests is implemented in Python (404 SLOC).
As discussed earlier, due to \craft's unavailable source code, we can only interpret its published artifacts~\cite{craftdroid_repo}; that functionality is implemented in Python (86 SLOC).
The data analyses that interpret our final datasets are written in R (585 SLOC).

\subsubsection{{Final Datasets}}
Our final datasets contain the results of 11,917 test transfer cases generated by the GUI Mappers from the existing techniques and our two baselines when applied on \framework's benchmark.
We apply 5 techniques---\appflow, ATM, GTM, Na\"ive, Perfect---to transfer  tests across  20 shopping and news apps, involving 1,000 source-target app pairs (2 app categories $\times$ 100 app pairs in each category $\times$ 5 techniques). This yielded 2,381 result entries per technique. 
As discussed earlier, we have to rely on \craft's final results, and can thus only compare \craft~ to the other  techniques on the 3 shopping apps---Geek, Wish, Etsy---used both in our benchmark and in \craft's evaluation. This gave us 12 result entries for \craft~ since only 2 tests are transferred by \craft~ in each app.
Each of the total 11,917 result entries contains the following information:
(1)  the source  and  target apps; 
(2)  the source, transferred, and ground-truth tests;
(3)  the technique used to transfer the test; 
(4)~the \emph{correct}/\emph{incorrect}/\emph{missed}/\emph{nonExist} sets of GUI events output by \framework's Fidelity Evaluator as described in Algorithm~\ref{alg:accuracy}, and the seven corresponding fidelity metrics  defined in Section~\ref{sec:design:metrics:accuracy}; and
(5) values of the two \emph{utility} metrics---\emph{effort} and \emph{reduction}---defined in Section~\ref{sec:design:metrics:utility}.
Note that obtaining these 11,917 result entries following prior work's evaluation processes would have required  manual inspection of 53,963 events that appear across all of the source tests, which is infeasible in practice.

\section{Findings}
\label{sec:results}

The datasets produced by \framework include the results obtained by evaluating side-by-side the extracted GUI Mapper components from the four existing test-reuse techniques \cite{hu2018appflow,behrang2018test,behrang2019atm,lin2019craftdroid} and the two baseline techniques we developed. In turn, this data enables further in-depth studies of a range of research questions in this emerging domain. 
As an illustration, this section highlights several findings uncovered by \framework's datasets that are missed by prior work. 

\subsection{GUI Mapper Comparison}
\label{sec:results:gui_mapper}

As discussed earlier, existing techniques are evaluated in their entirety, on different benchmark apps and tests, and using different evaluation metrics, all of which makes their results hard to compare.
By contrast, \framework was able to evaluate their extracted GUI Mappers side-by-side, with our two techniques---Na\"ive and Perfect---serving as baselines. 
We note that it is possible for a given test-reuse technique to produce results as a whole that may be different from those produced only by its extracted GUI Mapper. One reason may be that there is additional relevant functionality that is scattered across the technique's implementation. 
However, any such functionality can be added to the existing GUI Mappers, or  introduced in additional \framework components.

\subsubsection{{Fidelity Comparison}}
\label{sec:results:gui_mapper:accuracy}

\begin{figure}[t!]
	\centering
		\includegraphics[width=0.46\textwidth]{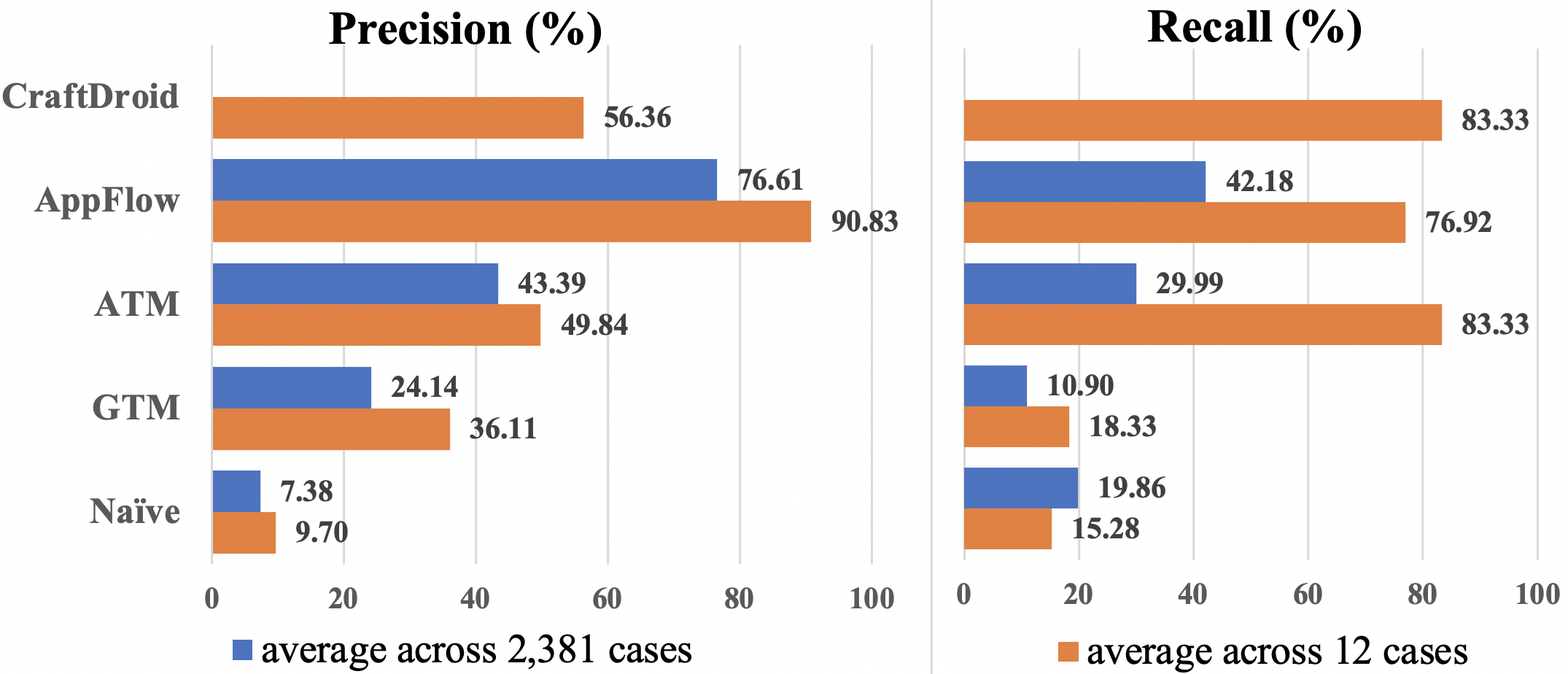}
	\caption{Comparison of average \emph{precision} and \emph{recall}.}
	\label{fig:accuracy}
\end{figure}

\framework's website~\cite{fruiterrepo} contains the results of all seven fidelity metrics from Section~\ref{sec:design:metrics:accuracy} obtained using our benchmark. 
Due to space limitations, we show the results of three metrics (Precision, Recall, Accuracy), and restrict our discussion to {Precision} and {Recall} since {Accuracy} follows a similar trend as {Precision}; the results of the four remaining fidelity metrics (Correct, Incorrect, Missed, NonExist) can provide an in-depth understanding on each of the four specific cases, and can be found on \framework's website \cite{fruiterrepo}. 
Figure~\ref{fig:accuracy} shows the average precision and recall achieved by the four existing techniques as well as Na\"ive; we omit Perfect since its values are always 100\% by definition.

\begin{figure}[t!]
	\centering
		\includegraphics[width=0.37\textwidth]{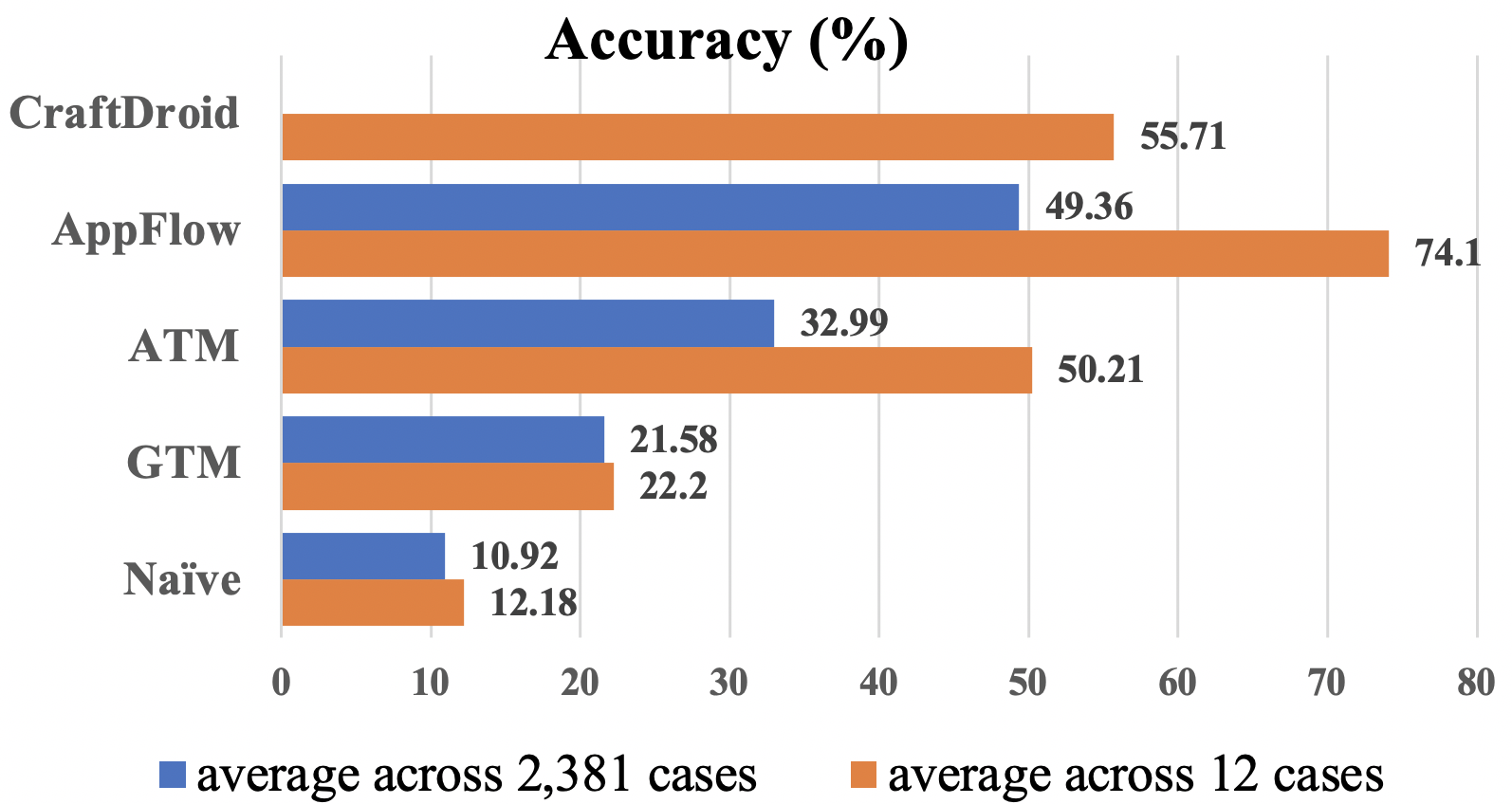}
	\caption{Comparison of average \emph{accuracy}.}
	\label{fig:accuracy_metric}
\end{figure}

For each technique except \craft, the top  (\textcolor{blue}{blue}) bar shows the {average} calculated based on 2,381 cases transferred among both shopping and news apps. 
To meaningfully compare \craft~ with other techniques, even if only partially, we  show the  {averages} calculated based on the 12 cases for which we have \craft's data, in the bottom  (\textcolor{orange}{orange}) bars.
\craft~ only transferred ``Sign In'' and ``Sign Up'' tests in the 3 shopping apps---Geek, Wish, and Etsy---leading to the 12 cases (6 source-target app pairs $\times$ 2 tests).

We highlight three  observations based on the results from Figure~\ref{fig:accuracy}. 
First, every existing technique yields lower recall than precision on the larger (\textcolor{blue}{blue}) data set, meaning that it suffers from more \emph{missed} (i.e., false negative) than \emph{incorrect} (i.e., false positive) cases. 
Although \appflow's recall is highest among the existing techniques, it exhibits the largest drop-off between its precision and recall values.  
A plausible  explanation is that, as an ML-based technique, \appflow~ will likely fail to recognize relevant GUI events if no similar events exist in its training data. This was somewhat unexpected, however, given that \appflow's authors carefully crafted its ML model to the app categories we also used in \framework, and suggests that additional research is needed in selecting and training effective ML models for UI test reuse.
By comparison, similarity-based techniques such as ATM will miss fewer GUI events in principle: they can always compute a similarity score between two  events and return the mapped events whose scores are above a given  threshold.
However, if the similarity threshold is set too low, it will 
 result in more \emph{incorrect} cases, leading to low precision.

A related observation is that \appflow's precision  outperforms the  other techniques across the board, for both the larger (\textcolor{blue}{blue}) and smaller (\textcolor{orange}{orange}) datasets. This is because \appflow~ has the advantage of  more information, obtained from a large corpus of apps  in its training dataset, than the similarity-based techniques, 
which compute the similarity scores based only on the information extracted from the source  and  target apps under analysis. However, \appflow's  recall is lower than both ATM and \craft~ on the 12 (\textcolor{orange}{orange}) cases from Geek, Wish, and Etsy. 
This reinforces the above observation that an ML-based technique will fail to recognize  GUI events if no similar events exist in its training data.

Finally, our data confirms that ATM  indeed improves upon GTM, as indicated in  their pairwise comparisons across both precision and recall, on both large and small datasets. 
In fact, GTM exhibits the lowest fidelity of all existing techniques, and its recall across the 2,381 (\textcolor{blue}{blue}) cases is actually lower than that achieved by the Na\"ive strategy.
We note that GTM's design is geared to transferring tests in programming assignments that share identical requirements, and is clearly not  suited to  heterogenous real-world apps. 

\subsubsection{{Utility Comparison}}
\label{sec:results:gui_mapper:utility}

Figure~\ref{fig:utility} shows the two utility metrics yielded by each of the four existing and two baselines. 
Recall from Section~\ref{sec:design:metrics:utility} that utility measures how useful the transferred tests are in practice compared to the ground-truth tests. The objective of  utility  is to minimize the \emph{effort} while maximizing the \emph{reduction}. 
 
The utility of existing techniques shows  similar trends to those observed in the case of fidelity. For example,  \appflow~ outperforms other  techniques, while GTM exhibits similar performance  to that of Na\"ive.
This indicates a possible correlation between the fidelity of the GUI event mapping and the utility of the transferred tests.

At the same time, we observe that, while our Perfect GUI Mapper achieves higher utility than the remaining techniques, that utility is not optimal.
In fact, Perfect's average \emph{reduction} is under 50\% across the 2,381 cases in the larger dataset (top, \textcolor{blue}{blue} bar). In other words, even with the best possible mapping strategy, we  save less than half of the effort required to complete the task manually. The previously published techniques perform much worse than this:  \appflow~saves under 30\%, ATM under 10\%, and GTM under 1\% of the required manual effort, while the reduction yielded by \craft~ on the smaller (\textcolor{orange}{orange}) dataset is lower than Perfect's on either of the two datasets. This indicates that fidelity is clearly not the only factor to  consider in order to achieve desired utility, and that there is large room for improvement in future test-reuse techniques.

To verify the above insights, we conducted pairwise correlation tests between the seven fidelity  and two utility metrics. 
Overall, the results, further discussed below and provided in their entirety in \framework's online repository~\cite{fruiterrepo}, show a \emph{weak correlation} between fidelity and utility. 
This reinforces our observation that accurate GUI  mappings can yield  useful transferred tests, but are not the only relevant factor. 
In turn, this  finding calls for exploration of other components in test-reuse techniques since the focus on GUI event mapping alone can hit a ``ceiling'', as shown by the Perfect baseline.
We discuss such possible directions next.

\subsection{Insights and Future Directions}
\label{sec:results:insights}

Guided by the above observations, 
we  explore potential strategies for improving UI test reuse with various statistical tests and manual inspections on \framework's datasets.
Due to space limitations, we highlight four  findings that were not reported by previous work.

\begin{figure}[t]
	\centering
		\includegraphics[width=0.46\textwidth]{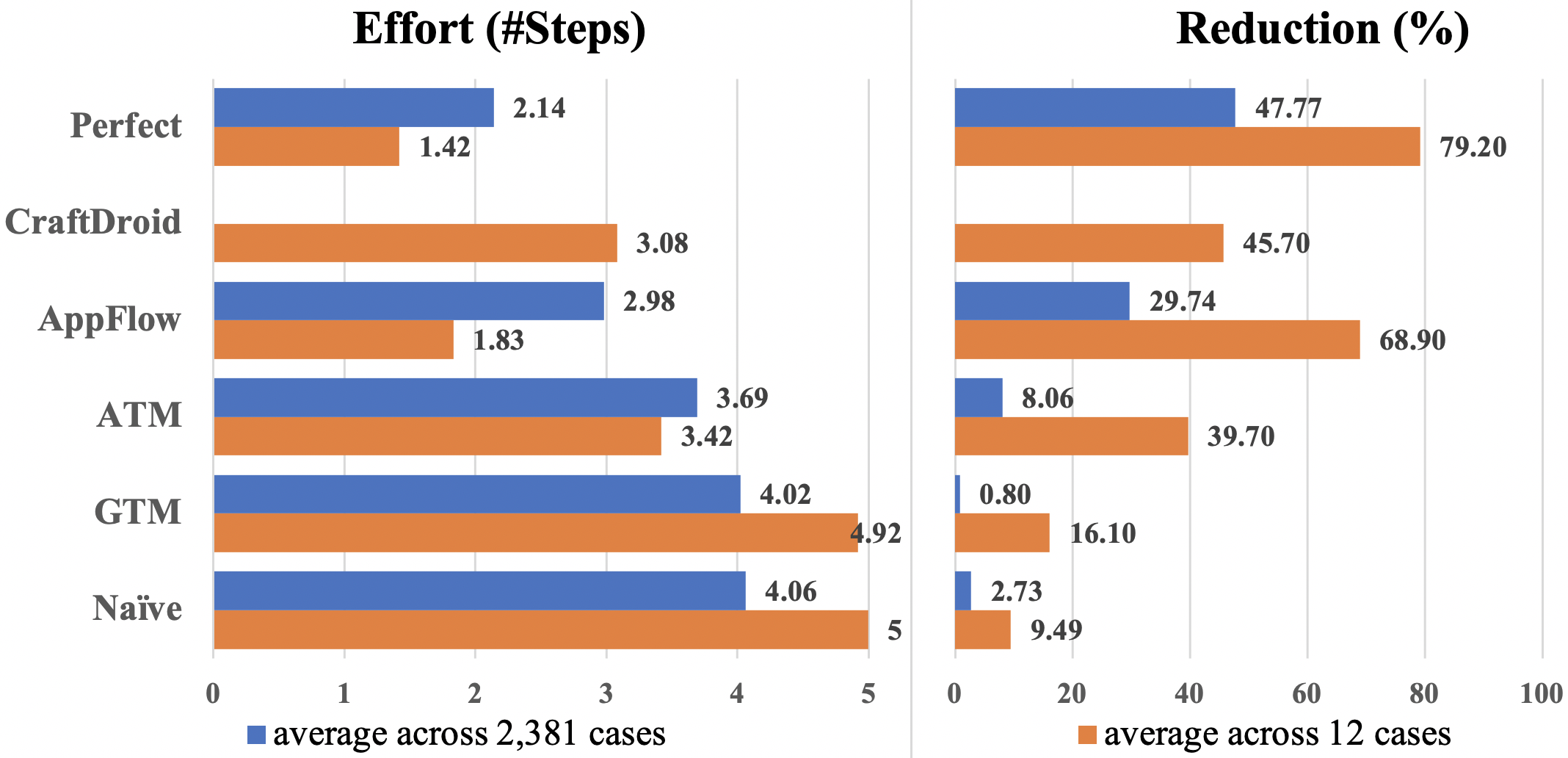}
	\caption{Comparison of average \emph{effort} and \emph{reduction}.}
	\label{fig:utility}
\end{figure}

\textbf{Source app selection matters for a given target app.} Figures~
\ref{fig:accuracy}, \ref{fig:accuracy_metric}, and \ref{fig:utility} all show consistent improvement across the techniques in the smaller datasets (12 cases transferred among 3 apps) compared to the larger ones (2,381 cases transferred among 20 apps). This suggests that certain source-target app pairs  achieve better results than others.
For example, we found that app pairs involving Wish, Geek, and a benchmark app called Home---all of which are developed by the same company, \emph{Wish Inc.}---
achieve high  fidelity and utility, regardless of the technique  used. 
Another such compatible app pair is ABC News and Reuters. Performing a large-scale evaluation enabled by \framework will help spot  pairings like this, and give researchers a starting point to explore the characteristics that can lead to better transfer results.

\textbf{Automated transfer is not suitable for  all tests.}
Our  utility metrics revealed large \emph{effort} and negative \emph{reduction} in some cases, meaning that correcting a transferred test required more work than  writing it from scratch.
Further inspection revealed that this is primarily due to a test's length rather than a technique's accuracy.
For instance, Perfect showed no benefit (\emph{reduction}~$\leq$ 0)~16\% of the time, and the average number of  source events in those cases is only 4.
This suggests that, for simple tests, manual construction may be preferable. 
Future research should consider the criteria for suitable tests to transfer instead of transferring all  source tests. 

\textbf{There is a trade-off between ML- and similarity-based techniques.}
As discussed above, an insufficient \emph{training set} in an ML-based technique may yield low recall, while a low  \emph{similarity threshold} in a similarity-based technique can address this but may yield low precision.
This suggests two future research directions. First, selecting training sets and similarity thresholds is important, but existing techniques did not justify their choices~\cite{hu2018appflow,behrang2018test,behrang2019atm,lin2019craftdroid}. 
There is clearly a need for further study of novel strategies such as incorporating dynamic selection criteria based on target app characteristics. 
Second, future research should consider the trade-offs across different test-reuse techniques and provide guidance on selecting the most suitable techniques for a given scenario.

\textbf{Test length is not a key factor influencing fidelity.}
\craft~ and GTM studied the relationship between the test length and their transferred results.
For instance, \craft~ showed a strong negative correlation between test length and its two fidelity metrics  (coefficient $<-0.5$ in both cases).
To verify these  findings, we conducted correlation tests on \framework's much larger datasets.
Our results indicate a negative but very weak correlation between test length and \framework's  fidelity metrics ($-0.25<$~coefficient~$<0$ across all seven cases).
This shows that test length is not the key factor that impacts fidelity, arguing that future research targeting reuse of complex tests may be a fruitful direction.

\section{Conclusion}
\label{sec:conclusion}
\vspace{-1mm}

This paper has presented \framework, a customizable framework for automatically evaluating UI test-reuse techniques. \framework~ has been instantiated and successfully demonstrated on the key functionality extracted from existing test-reuse techniques that target Android apps. In the process, we have been able to identify several avenues of future research that prior work has either missed or actually flagged as not viable. We publicly release \framework, its accompanying artifacts, and all of our evaluation data, as a way of fostering future research in this area of growing interest and importance.
\begin{acks}
We would like to thank Farnaz Behrang, Gang Hu, and Jun-Wei Lin, for their generous help on explaining the details of their respective test-reuse techniques to us. 
This work is supported by the U.S. National Science Foundation under grants 1717963 and 1823354, U.S. Office of Naval Research under grant N00014-17-1-2896, and ERC Advanced Fellowship Grant no. 741278 (EPIC).
\end{acks}

\clearpage
\bibliographystyle{ACM-Reference-Format}
\bibliography{FSE2020-FrUITeR}


\begin{thebibliography}{28}


\ifx \showCODEN    \undefined \def \showCODEN     #1{\unskip}     \fi
\ifx \showDOI      \undefined \def \showDOI       #1{#1}\fi
\ifx \showISBNx    \undefined \def \showISBNx     #1{\unskip}     \fi
\ifx \showISBNxiii \undefined \def \showISBNxiii  #1{\unskip}     \fi
\ifx \showISSN     \undefined \def \showISSN      #1{\unskip}     \fi
\ifx \showLCCN     \undefined \def \showLCCN      #1{\unskip}     \fi
\ifx \shownote     \undefined \def \shownote      #1{#1}          \fi
\ifx \showarticletitle \undefined \def \showarticletitle #1{#1}   \fi
\ifx \showURL      \undefined \def \showURL       {\relax}        \fi
\providecommand\bibfield[2]{#2}
\providecommand\bibinfo[2]{#2}
\providecommand\natexlab[1]{#1}
\providecommand\showeprint[2][]{arXiv:#2}

\bibitem[\protect\citeauthoryear{??}{app}{2019a}]%
        {appflowrepo}
 \bibinfo{year}{2019}\natexlab{a}.
\newblock \bibinfo{title}{AppFlow's source code and artifacts}.
\newblock \bibinfo{howpublished}{\url{https://github.com/columbia/appflow}}.
\newblock


\bibitem[\protect\citeauthoryear{??}{app}{2019b}]%
        {appium}
 \bibinfo{year}{2019}\natexlab{b}.
\newblock \bibinfo{title}{{Appium: Mobile App Automation Made Awesome.}}
\newblock
\newblock
\urldef\tempurl%
\url{http://appium.io}
\showURL{%
\tempurl}


\bibitem[\protect\citeauthoryear{??}{atm}{2019}]%
        {atm_repo}
 \bibinfo{year}{2019}\natexlab{}.
\newblock \bibinfo{title}{ATM's source code and artifacts}.
\newblock
  \bibinfo{howpublished}{\url{https://sites.google.com/view/apptestmigrator/}}.
\newblock


\bibitem[\protect\citeauthoryear{??}{cli}{2019}]%
        {click}
 \bibinfo{year}{2019}\natexlab{}.
\newblock \bibinfo{title}{{Click - Appium}}.
\newblock
\newblock
\urldef\tempurl%
\url{http://appium.io/docs/en/commands/element/actions/click}
\showURL{%
\tempurl}


\bibitem[\protect\citeauthoryear{??}{cra}{2019}]%
        {craftdroid_repo}
 \bibinfo{year}{2019}\natexlab{}.
\newblock \bibinfo{title}{CraftDroid's source code and artifacts}.
\newblock
  \bibinfo{howpublished}{\url{https://sites.google.com/view/craftdroid/}}.
\newblock


\bibitem[\protect\citeauthoryear{??}{esp}{2019}]%
        {espresso}
 \bibinfo{year}{2019}\natexlab{}.
\newblock \bibinfo{title}{Espresso}.
\newblock
  \bibinfo{howpublished}{\url{https://developer.android.com/training/testing/espresso}}.
\newblock


\bibitem[\protect\citeauthoryear{??}{and}{2019}]%
        {android_find_elements}
 \bibinfo{year}{2019}\natexlab{}.
\newblock \bibinfo{title}{{Find Elements - Appium}}.
\newblock
\newblock
\urldef\tempurl%
\url{http://appium.io/docs/en/commands/element/find-elements}
\showURL{%
\tempurl}


\bibitem[\protect\citeauthoryear{??}{gtm}{2019}]%
        {gtm_repo}
 \bibinfo{year}{2019}\natexlab{}.
\newblock \bibinfo{title}{GTM's source code and artifacts}.
\newblock
  \bibinfo{howpublished}{\url{https://sites.google.com/view/testmigration/}}.
\newblock


\bibitem[\protect\citeauthoryear{??}{And}{2019}]%
        {AndroidActivity}
 \bibinfo{year}{2019}\natexlab{}.
\newblock \bibinfo{title}{{Introduction to Activities {$\vert$} Android
  Developers}}.
\newblock
\newblock
\urldef\tempurl%
\url{https://developer.android.com/guide/components/activities/intro-activities}
\showURL{%
\tempurl}


\bibitem[\protect\citeauthoryear{??}{sen}{2019}]%
        {sendkeys}
 \bibinfo{year}{2019}\natexlab{}.
\newblock \bibinfo{title}{{Send Keys - Appium}}.
\newblock
\newblock
\urldef\tempurl%
\url{http://appium.io/docs/en/commands/element/actions/send-keys}
\showURL{%
\tempurl}


\bibitem[\protect\citeauthoryear{??}{soo}{2019}]%
        {soot}
 \bibinfo{year}{2019}\natexlab{}.
\newblock \bibinfo{title}{Soot - A Java optimization framework}.
\newblock \bibinfo{howpublished}{\url{https://github.com/Sable/soot}}.
\newblock


\bibitem[\protect\citeauthoryear{??}{fru}{2020}]%
        {fruiterrepo}
 \bibinfo{year}{2020}\natexlab{}.
\newblock \bibinfo{title}{FrUITeR's website}.
\newblock \bibinfo{howpublished}{\url{https://felicitia.github.io/FrUITeR/}}.
\newblock


\bibitem[\protect\citeauthoryear{??}{Ghe}{2020}]%
        {Gherkin}
 \bibinfo{year}{2020}\natexlab{}.
\newblock \bibinfo{title}{{Gherkin Syntax - Cucumber Documentation}}.
\newblock
\newblock
\urldef\tempurl%
\url{https://cucumber.io/docs/gherkin}
\showURL{%
\tempurl}


\bibitem[\protect\citeauthoryear{??}{web}{2020a}]%
        {webidxpath}
 \bibinfo{year}{2020}\natexlab{a}.
\newblock \bibinfo{title}{{How to locate an element on the page - Web
  Performance}}.
\newblock
\newblock
\urldef\tempurl%
\url{https://www.webperformance.com/load-testing-tools/blog/articles/real-browser-manual/building-a-testcase/how-locate-element-the-page}
\showURL{%
\tempurl}


\bibitem[\protect\citeauthoryear{??}{jso}{2020}]%
        {json}
 \bibinfo{year}{2020}\natexlab{}.
\newblock \bibinfo{title}{JSON - Wikipedia}.
\newblock
\newblock
\urldef\tempurl%
\url{https://en.wikipedia.org/wiki/JSON}
\showURL{%
\tempurl}


\bibitem[\protect\citeauthoryear{??}{sel}{2020}]%
        {selenium}
 \bibinfo{year}{2020}\natexlab{}.
\newblock \bibinfo{title}{{SeleniumHQ Browser Automation}}.
\newblock
\newblock
\urldef\tempurl%
\url{https://www.selenium.dev}
\showURL{%
\tempurl}


\bibitem[\protect\citeauthoryear{??}{web}{2020b}]%
        {webelement}
 \bibinfo{year}{2020}\natexlab{b}.
\newblock \bibinfo{title}{{Web element :: Documentation for Selenium}}.
\newblock
\newblock
\urldef\tempurl%
\url{https://selenium.dev/documentation/en/webdriver/web_element}
\showURL{%
\tempurl}


\bibitem[\protect\citeauthoryear{admin}{admin}{2019}]%
        {find_element_id}
\bibfield{author}{\bibinfo{person}{admin}.} \bibinfo{year}{2019}\natexlab{}.
\newblock \showarticletitle{{Chapter-4: Appium Locator Finding Strategies -
  Kobiton}}.
\newblock \bibinfo{journal}{\emph{Kobiton}} (\bibinfo{date}{Apr}
  \bibinfo{year}{2019}).
\newblock
\urldef\tempurl%
\url{https://kobiton.com/book/chapter-4-appium-locator-finding-strategies}
\showURL{%
\tempurl}


\bibitem[\protect\citeauthoryear{Allen and Cocke}{Allen and Cocke}{1976}]%
        {allen1976program}
\bibfield{author}{\bibinfo{person}{Frances~E. Allen} {and}
  \bibinfo{person}{John Cocke}.} \bibinfo{year}{1976}\natexlab{}.
\newblock \showarticletitle{A program data flow analysis procedure}.
\newblock \bibinfo{journal}{\emph{Commun. ACM}} \bibinfo{volume}{19},
  \bibinfo{number}{3} (\bibinfo{year}{1976}), \bibinfo{pages}{137}.
\newblock


\bibitem[\protect\citeauthoryear{Behrang and Orso}{Behrang and Orso}{2018}]%
        {behrang2018test}
\bibfield{author}{\bibinfo{person}{Farnaz Behrang} {and}
  \bibinfo{person}{Alessandro Orso}.} \bibinfo{year}{2018}\natexlab{}.
\newblock \showarticletitle{Test migration for efficient large-scale assessment
  of mobile app coding assignments}. In \bibinfo{booktitle}{\emph{Proceedings
  of the 27th ACM SIGSOFT International Symposium on Software Testing and
  Analysis}}.
\newblock


\bibitem[\protect\citeauthoryear{Behrang and Orso}{Behrang and Orso}{2019}]%
        {behrang2019atm}
\bibfield{author}{\bibinfo{person}{Farnaz Behrang} {and}
  \bibinfo{person}{Alessandro Orso}.} \bibinfo{year}{2019}\natexlab{}.
\newblock \showarticletitle{Test Migration Between Mobile Apps with Similar
  Functionality}. In \bibinfo{booktitle}{\emph{34th International Conference on
  Automated Software Engineering (ASE 2019)}}.
\newblock


\bibitem[\protect\citeauthoryear{Choudhary, Gorla, and Orso}{Choudhary
  et~al\mbox{.}}{2015}]%
        {choudhary2015orsosurvey}
\bibfield{author}{\bibinfo{person}{Shauvik~Roy Choudhary},
  \bibinfo{person}{Alessandra Gorla}, {and} \bibinfo{person}{Alessandro Orso}.}
  \bibinfo{year}{2015}\natexlab{}.
\newblock \showarticletitle{Automated test input generation for android: Are we
  there yet? (E)}. In \bibinfo{booktitle}{\emph{2015 30th IEEE/ACM
  International Conference on Automated Software Engineering (ASE)}}.
\newblock


\bibitem[\protect\citeauthoryear{Hu, Zhu, and Yang}{Hu et~al\mbox{.}}{2018}]%
        {hu2018appflow}
\bibfield{author}{\bibinfo{person}{Gang Hu}, \bibinfo{person}{Linjie Zhu},
  {and} \bibinfo{person}{Junfeng Yang}.} \bibinfo{year}{2018}\natexlab{}.
\newblock \showarticletitle{AppFlow: using machine learning to synthesize
  robust, reusable UI tests}. In \bibinfo{booktitle}{\emph{Proceedings of the
  2018 26th ACM Joint Meeting on European Software Engineering Conference and
  Symposium on the Foundations of Software Engineering}}. ACM,
  \bibinfo{pages}{269--282}.
\newblock


\bibitem[\protect\citeauthoryear{Levenshtein}{Levenshtein}{1966}]%
        {levenshtein1966binary}
\bibfield{author}{\bibinfo{person}{Vladimir~I Levenshtein}.}
  \bibinfo{year}{1966}\natexlab{}.
\newblock \showarticletitle{Binary codes capable of correcting deletions,
  insertions, and reversals}. In \bibinfo{booktitle}{\emph{Soviet physics
  doklady}}, Vol.~\bibinfo{volume}{10}. \bibinfo{pages}{707--710}.
\newblock


\bibitem[\protect\citeauthoryear{Lin, Jabbarvand, and Malek}{Lin
  et~al\mbox{.}}{2019}]%
        {lin2019craftdroid}
\bibfield{author}{\bibinfo{person}{Jun-Wei Lin}, \bibinfo{person}{Reyhaneh
  Jabbarvand}, {and} \bibinfo{person}{Sam Malek}.}
  \bibinfo{year}{2019}\natexlab{}.
\newblock \showarticletitle{Test Transfer Across Mobile Apps Through Semantic
  Mapping}. In \bibinfo{booktitle}{\emph{34th International Conference on
  Automated Software Engineering (ASE 2019)}}.
\newblock


\bibitem[\protect\citeauthoryear{Linares-V{\'a}squez, Bernal-C{\'a}rdenas,
  Moran, and Poshyvanyk}{Linares-V{\'a}squez et~al\mbox{.}}{2017}]%
        {linares2017developers}
\bibfield{author}{\bibinfo{person}{Mario Linares-V{\'a}squez},
  \bibinfo{person}{Carlos Bernal-C{\'a}rdenas}, \bibinfo{person}{Kevin Moran},
  {and} \bibinfo{person}{Denys Poshyvanyk}.} \bibinfo{year}{2017}\natexlab{}.
\newblock \showarticletitle{How do developers test android applications?}. In
  \bibinfo{booktitle}{\emph{2017 IEEE International Conference on Software
  Maintenance and Evolution (ICSME)}}.
\newblock


\bibitem[\protect\citeauthoryear{Manning, Raghavan, and Sch{\"u}tze}{Manning
  et~al\mbox{.}}{2008}]%
        {accuracy_precision_recall}
\bibfield{author}{\bibinfo{person}{Christopher~D Manning},
  \bibinfo{person}{Prabhakar Raghavan}, {and} \bibinfo{person}{Hinrich
  Sch{\"u}tze}.} \bibinfo{year}{2008}\natexlab{}.
\newblock \bibinfo{booktitle}{\emph{Introduction to information retrieval}}.
\newblock \bibinfo{publisher}{Cambridge university press}.
\newblock


\bibitem[\protect\citeauthoryear{Rau, Hotzkow, and Zeller}{Rau
  et~al\mbox{.}}{2018}]%
        {zeller2018transferring}
\bibfield{author}{\bibinfo{person}{Andreas Rau}, \bibinfo{person}{Jenny
  Hotzkow}, {and} \bibinfo{person}{Andreas Zeller}.}
  \bibinfo{year}{2018}\natexlab{}.
\newblock \showarticletitle{Transferring Tests Across Web Applications}. In
  \bibinfo{booktitle}{\emph{Web Engineering}},
  \bibfield{editor}{\bibinfo{person}{Tommi Mikkonen}, \bibinfo{person}{Ralf
  Klamma}, {and} \bibinfo{person}{Juan Hern{\'a}ndez}} (Eds.).
  \bibinfo{publisher}{Springer International Publishing},
  \bibinfo{address}{Cham}, \bibinfo{pages}{50--64}.
\newblock
\showISBNx{978-3-319-91662-0}


\end{thebibliography}
\end{document}